\documentclass[sn-standardnature,onecolumn]{sn-jnl}%

\jyear{2022}%

\theoremstyle{thmstyleone}%
\theoremstyle{thmstyletwo}%

\theoremstyle{thmstylethree}%

\usepackage{placeins}
\begin{document}

\title[Hysteresis in SWCC]{Investigating the source of hysteresis in the Soil-Water Characteristic Curve using the multiphase lattice Boltzmann method}


\author*[1]{\fnm{Reihaneh} \sur{Hosseini}}\email{reihos@utexas.edu}

\author[1]{\fnm{Krishna} \sur{Kumar}}

\author[2]{\fnm{Jean-Yves} \sur{Delenne}}

\affil*[1]{\orgdiv{Department of Civil, Architectural, and Environmental Engineering}, \orgname{the University of Texas at Austin}, \orgaddress{\street{301 E Dean Keaton}, \city{Austin}, \postcode{78731}, \state{Texas}, \country{USA}}}

\affil[2]{\orgdiv{IATE, Université Montpellier, CIRAD, INRAE, Montpellier SupAgro},  \orgaddress{\city{Montpellier}, \country{France}}}



\abstract{The soil-water characteristic curve (SWCC) is the most fundamental relationship in unsaturated soil mechanics, relating the amount of water in the soil to the corresponding matric suction. From experimental evidence, it is known that SWCC exhibits hysteresis (i.e. wetting/drying path dependence). Various factors have been proposed as contributors to SWCC hysteresis, including air entrapment, contact angle hysteresis, ink-bottle effect, and change of soil fabric due to swelling and shrinkage, however, the significance of their contribution is debated. From our pore-scale numerical simulations, using the multiphase lattice Boltzmann method, we see that even when controlling for all these factors SWCC hysteresis still occurs, indicating that there is some underlying source that is not accounted for in these factors. We find this underlying source by comparing the liquid/gas phase distributions for simulated wetting and drying experiments of 2D and 3D granular packings. We see that during wetting (i.e. pore filling) many liquid bridges expand simultaneously and join together to fill the pores from the smallest to the largest, allowing menisci with larger radii of curvature (lower matric suction). Whereas, during drying (i.e. pore emptying), only the limited existing gas clusters can expand, which become constrained by the size of the pore openings surrounding them and result in menisci with smaller radii of curvature (higher matric suction).}




\maketitle

\section{Introduction}\label{sec1}

The most fundamental relationship in unsaturated soil mechanics is the relation between the amount of water in the soil, usually expressed as volumetric water content or degree of saturation, and its matric suction, defined as the difference between the pore air and pore water pressures \cite{Fredlund2012,lu2004unsaturated}. This relationship is commonly known as the soil-water retention curve (SWRC) or the soil-water characteristic curve (SWCC). In geotechnical engineering, we use empirical and theoretical models of SWCC to relate easy-to-measure soil index properties (e.g., volumetric water content) with harder-to-measure matric suction in unsaturated soils. The SWCC is commonly used in construction and operation of earth dams, stability analysis of natural slopes, design of soil covers, and predictions of shrinkage and swelling of expansive soils \cite{Fredlund2012}.

It is well known that the SWCC for a given soil is not a one-to-one function and depends on the wetting and drying history of the soil, however, the source of this hysteresis is not well understood \cite{lu2004unsaturated}. The main theories that have been proposed for the cause of SWCC hysteresis include air entrapment, contact angle hysteresis, ink-bottle effect, and change of soil fabric due to swelling and shrinkage. Air entrapment refers to the phenomena where air bubbles are trapped at dead-end pores with the water bypassing them \cite{Bear1979,lu2004unsaturated}. While there is experimental evidence suggesting that air-entrapment occurs during wetting \cite{Poulovassilis1970,Fredlund2012}, it has not been proven to be the main source of hysteresis. Contact angle hysteresis refers to the difference between the wetting and drying soil-water contact angle. Contact angle hysteresis has been proven theoretically and experimentally only for a single liquid droplet or liquid bridge \cite{Gao2006,DeSouza2008}, and therefore, has only been associated with SWCC hysteresis at low water contents where the liquid is only in the form of bridges between grains \cite{Likos2004}. The ink-bottle effect has been attributed to the case where the pore structure is non-homogeneous, consisting of a series of narrower pores connected to wider pores. It has been suggested that such pore structure can hold more water during drying compared to wetting for the same suction level. This phenomenon has only been illustrated schematically in the literature \cite{Haines1930,Bear1979,lu2004unsaturated} and has not been shown experimentally. Finally, change of soil fabric has been shown experimentally to affect SWCC hysteresis \cite{Estabragh2015}, however, it only applies to fine-grained soils and does not explain the hysteresis observed for coarse-grained soils. 

More recently, with the advancements in X-ray computed tomography imaging, a number of studies have focused on identifying the source of SWCC hysteresis by visualizing the liquid and gas phase distribution inside a small unsaturated soil sample during wetting and drying \cite{Higo2015,Kido2020}. While, in these studies, differences between the liquid and gas phase distributions on the wetting and drying paths have been identified, the source of SWCC hysteresis is still unclear. Identifying the source of hysteresis can have a great impact on enhancing the SWCC models used in geotechnical engineering applications. 

In this study, we investigate the source of SWCC hysteresis by means of numerical simulations at the pore scale. The advantages of using pore-scale numerical modeling for this purpose include: 1) the ability to start with a very simple idealized model and step-by-step build to more complex models, to delineate individual effects, 2) monitor the liquid and gas phase distributions at pore scale to find where the difference in the distributions during drying and wetting arise from, and 3) systematically control for the potential causes of SWCC hysteresis proposed in the literature, to find the underlying cause. For the latter, we implement the following strategies. We ensure that the contact angle at equilibrium is constant during both drying and wetting, therefore, eliminating contact angle hysteresis as a potential source of SWCC hysteresis. Next, rather than draining/injecting liquid from the boundary of the soil sample, we use a liquid drainage/injection scheme that decreases/increases the amount of liquid everywhere in the system simultaneously, similar to placing the soil sample in a humidity chamber. This way we eliminate the ink-bottle effect, because it supposedly occurs when the liquid flows in the exact opposite directions during drying versus wetting, as well as air entrapment. Finally, we keep the grains fixed in our models, thereby, eliminating the effects of soil fabric change.

We use the multiphase lattice Boltzmann method for fluid simulation at pore-scale. We discuss the details of this method in Section \ref{sec2}. In Section \ref{sec3}, we simulate a 2D liquid bridge between two solid plates, to show the origin of matric suction in unsaturated material and to reveal that for a liquid bridge between planar surfaces suction is not a function of degree of saturation. In this section we also measure the surface tension and contact angle, and show that if the model is allowed to reach equilibrium at each step of wetting or drying, the contact angle at equilibrium remains constant. In Section \ref{sec4}, we simulate a 2D bridge between two circular grains, to show how suction changes with degree of saturation for a liquid bridge between non-planar solid surfaces. We also verify that there is no wetting/drying hysteresis for the case of a liquid bridge between two grains. In Section \ref{sec5}, we simulate drying and wetting experiments for a synthetic 2D granular model and find the source of hysteresis by comparing the liquid/gas phase distributions on the two paths. Finally we present a synthetic 3D granular model, which could be thought of as a small sample of rounded sand, to show that the conclusions made for the source of SWCC in 2D also apply to 3D.

\section{Numerical method}\label{sec2}

The multiphase lattice Boltzmann method (LBM) which is an extension to the single-phase LBM, is a numerical method that allows simulating fluid with multiple phases (i.e. liquid and gas) or multiple components (e.g. different types of liquids) at the pore-scale \cite{huang2015multiphase,Kruger2017}. Many studies have shown the potential of multiphase LBM for studying pore-scale mechanisms \cite{Pan2004, Schaap2007, Sukop2008, Pot2015}, however, only a few have investigated these mechanisms \cite{Delenne2015,Richefeu2016,Li2018}, and none have looked into hysteresis effects. 

\subsection{Single-phase LBM}\label{subsec1}

\begin{figure}[t]
\centering
\includegraphics[width=0.5\textwidth]{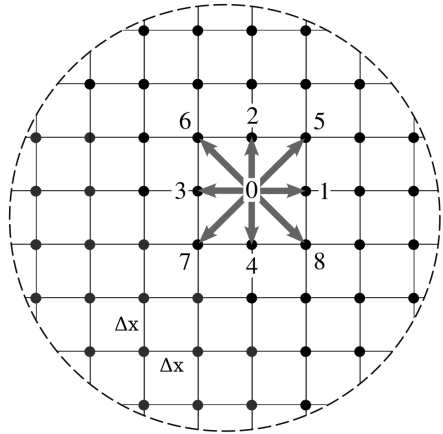}
\caption{An example LBM lattice and the D2Q9 velocity model. Numbers 0 to 8 correspond to the discrete velocity directions. The 0 direction refers to remaining in place.}
\label{lbm_grid}
\end{figure}

To simulate a fluid domain using LBM, the domain needs to be discretized with a grid of equal spacings, $\Delta x$, in every direction. This grid is called the lattice and the grid intersections are called lattice nodes. An example lattice for a small 2D subdomain is shown in Figure \ref{lbm_grid}. Each lattice node represents a collection of fluid molecules (particles), hence LBM simulates the fluid at the mesoscale \cite{Kruger2017}. At each lattice node, particles can only take discrete velocities, $\boldsymbol{c}_i$, where $i$ denotes the direction of the velocity, shown with gray arrows in Figure \ref{lbm_grid} for a 2D model. There are different ways of discretizing the velocity directions; we use the D2Q9 set \cite{Chen1989} for our 2D simulations (see Figure \ref{lbm_grid}), and the D3Q19 set \cite{DHumieres1986} for our 3D simulations, where the numbers following D and Q represent the dimension of the lattice and the number of discrete velocities, respectively \cite{Qian1992}. In all velocity sets, the magnitudes of the discrete velocities are chosen such that the particles can only travel to their neighboring nodes during one timestep of the simulation, $\Delta t$. Each lattice node at each discrete direction, $i$, has a property called the particle distribution function, $f_i$, which represents the density of the particles with velocity $\boldsymbol{c}_i$ at the given node. Macroscopic quantities such as mass density, $\rho$, or fluid velocity, $\boldsymbol{u}$, which are usually the parameters of interest in fluid dynamics, can be calculated from the distribution functions using
\begin{equation}
\rho(\boldsymbol{x},t)=\sum_if_i(\boldsymbol{x},t),
\end{equation}
and
\begin{equation}
\boldsymbol{u}(\boldsymbol{x},t)=\frac{\sum_i\boldsymbol{c}_if_i(\boldsymbol{x},t)}{\rho(\boldsymbol{x},t)}.
\end{equation}

\begin{figure*}[!ht]
\centering
\includegraphics[width=1\textwidth]{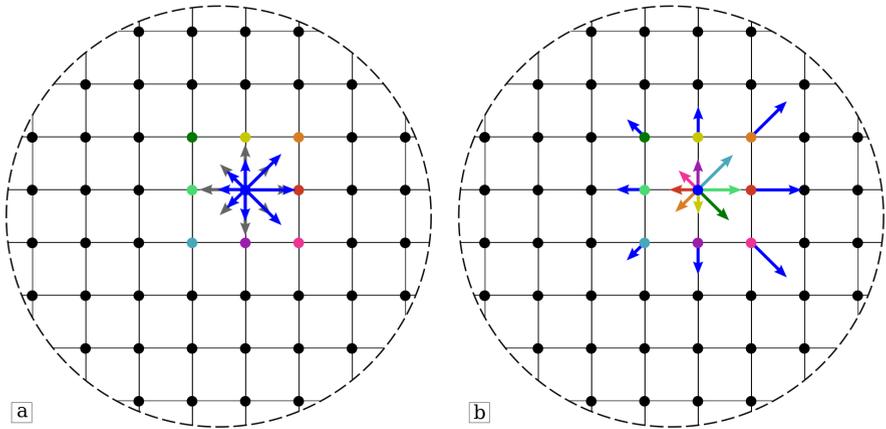}
\caption{The two steps of the LBM updating scheme: a) collision and b) streaming, shown for a single node. The arrows correspond to the distribution functions. Gray arrows are before collision, blue arrows are after collision, and other color arrows are after streaming, all for the single blue lattice node.
}
\label{collision_streaming}
\end{figure*}

The entire LBM simulation consists of updating the distribution functions at every $\Delta t$, until the equilibrium distribution functions are reached. The update process is usually done in two steps: collision and streaming. During the collision step, the distribution functions at each node are updated in place. In Figure \ref{collision_streaming}a, the gray arrows represent $f_i$ magnitudes for a single node before collision and the blue arrows represent the updated $f_i$ magnitudes for that node after collision; the same process is done at all other lattice nodes. Following collision, the updated distribution functions are streamed according to their velocities: each $f_i$ is moved one lattice node in the $i$ direction. In Figure \ref{collision_streaming}b, the blue node streams its updated distribution functions, which are shown with blue arrows, to its neighboring nodes, and, at the same time, receives the streamed distribution functions from its neighbors. Collision and streaming can be shown mathematically as
\begin{equation}
f_i(\boldsymbol{x}+\boldsymbol{c}_i\Delta t, t + \Delta t) = f_i(\boldsymbol{x}, t) + \Omega_i(\boldsymbol{x}, t),
\label{lbe}
\end{equation}
which is the discretized version of the Boltzmann equation \cite{He1997}, therefore known as the lattice Boltzmann equation. In Eq \ref{lbe}, $\Omega_i$ is the collision operator which determines the evolution of the distribution functions towards equilibrium. There are different models for the collision operator; we use the Bhatnagar, Gross and Krook \cite{bgk} operator, 
\begin{equation}
\Omega_i = -\frac{f_i - f_i^{eq}}{\tau}\Delta t,
\label{bgk}
\end{equation}
which relies on a single relaxation time, $\tau$. We use $\tau = \Delta t$ in our simulations. The equilibrium distribution function, $f_i^{eq}$, in Eq \ref{bgk} is given by 
\begin{equation}
f_i^{eq}(\boldsymbol{x}, t)=w_i\rho\left(1+\frac{\boldsymbol{u}.\boldsymbol{c}_i}{c_s^2}+
\frac{(\boldsymbol{u}.\boldsymbol{c}_i)^2}{2c_s^4}-
\frac{\boldsymbol{u}.\boldsymbol{u}}{2c_s^2}\right),
\label{feq}
\end{equation}
where $w_i$ are weights associated with the velocity set used and $c_s$ is the isothermal speed of sound, given as $\frac{1}{\sqrt{3}}\frac{\Delta x}{\Delta t}$ for the velocity sets used in this work (see \cite{He1997} for the full derivation of $f_i^{eq}$). 

Through the Chapman-Enskog analysis, the lattice Boltzmann equation can be linked to the Navier-Stokes equations \cite{Kruger2017}, and it can be shown that the kinematic shear viscosity in LBM is given by
\begin{equation}
\nu = c_s^2(\tau - \frac{\Delta t}{2}).
\end{equation}

Commonly, in an LBM simulation, $\Delta x$, $\Delta t$, and the average fluid density, $\rho_0$, are set to unity for simplicity. As a result, all other parameters, such as velocity, viscosity and pressure, will have normalized units called lattice units (lu). Lattice units can be converted to physical units by choosing proper conversion factors for space, time and mass density. The choice of these conversion factors, which are essentially $\Delta x$, $\Delta t$, and $\rho_0$ in physical units, should be based on the accuracy, stability, and efficiency of the simulation \cite{Kruger2017}. In addition, depending on the problem being simulated, the proper law of similarity, such as matching Reynolds number, Weber number or Capillary number, should be satisfied. Since in this study we are only interested in the qualitative behavior of the fluid phases in an unsaturated granular packing rather than exact quantities or comparison to a physical model, we present our results in lattice units.

\subsection{Multiphase LBM}\label{subsec2}

There are several different methods available to incorporate multiphase flow in LBM \cite{huang2015multiphase}; we use the popular Shan-Chen method \cite{Shan1993,Shan1994}. In the Shan-Chen method, the velocity, $\boldsymbol{u}$, used to calculate the equilibrium distribution function in Eq \ref{feq}, is modified to $\boldsymbol{u}^{eq}$ defined as
\begin{equation}
\boldsymbol{u}^{eq}=\frac{\sum_i\boldsymbol{c}_if_i+\tau\boldsymbol{F}_{SC}}{\rho},
\label{ueq}
\end{equation}
which includes an additional force term, $\boldsymbol{F}_{SC}$. This additional force term causes the fluid to separate into a phase with higher density and a phase with lower density. $\boldsymbol{F}_{SC}$ for a single-component fluid is defined by
\begin{equation}
\boldsymbol{F}_{SC}(\boldsymbol{x})=-\psi(\boldsymbol{x})G\sum w_i\psi(\boldsymbol{x}+\boldsymbol{c}_i\Delta t)\boldsymbol{c}_i\Delta t,
\label{fsc}
\end{equation}
where G is a scalar that controls the strength of the interaction between fluid elements (negative for attraction), and  $\psi(\boldsymbol{x})$ is a parameter controlled by the density of the fluid, $\psi(\boldsymbol{x})=\psi[\rho(\boldsymbol{x})]=\psi(\rho)$. Shan and Chen \cite{Shan1993} originally proposed $\psi(\rho) = \rho_0[1-\exp(-\rho/\rho_0)]$, however, we used an alternative form discussed below.

\begin{figure}[!htbp]
\centering
\includegraphics[width=0.5\textwidth]{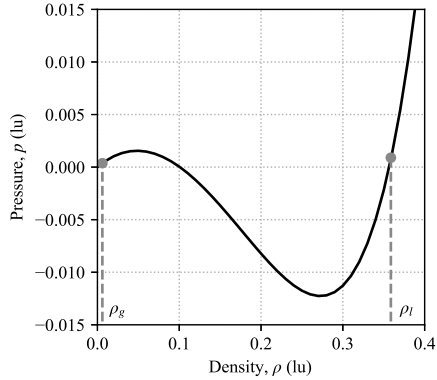}
\caption{The Carnahan-Starling EOS with $a=1$ lu, $b=4$ lu, and $R=1$ lu and $T=0.7Tc$. $\rho_l$ and $\rho_g$ are liquid and gas coexistence densities.}
\label{cs_eos_fig}
\end{figure}

The equation of state (EOS) for the Shan-Chen multiphase LBM model can be derived \cite{Benzi2006} as 
\begin{equation}
p(\rho)=c_s^2\rho+\frac{c_s^2\Delta t^2 G}{2}\psi^2(\rho),
\label{eos}
\end{equation}
which is essentially the isothermal EOS with an additional term that allows coexistence of liquid and gas phases. Yuan and Schaefer \cite{Yuan2006} showed that by modifying $\psi(\rho)$ any other EOS can be incorporated into the model. This is done by rearranging Eq \ref{eos} as 
\begin{equation}
\psi(\rho)=\sqrt{(p(\rho)-c_s^2\rho)\frac{2}{c_s^2\Delta t^2 G}},
\label{psi}
\end{equation}
and replacing $p(\rho)$ with the formulation of the particular EOS. We choose the Carnahan-Starling (C-S) EOS, which has been shown to have a better performance for high density ratios \cite{Yuan2006}. The C-S EOS is given by
\begin{equation}
p(\rho)=\rho RT\frac{1+b\rho/4+(b\rho/4)^2-(b\rho/4)^3}{(1-b\rho/4)^3}-a\rho^2,
\label{cs_eos}
\end{equation}
where $R$ is the gas constant, $T$ is temperature, $a$ is the attraction parameter, and $b$ is the repulsion parameter. It can be shown that $a = \frac{0.4963R^2T_c^2}{p_c}$ and $b =\frac{0.18727RT_c}{p_c}$, where $T_c$ is the temperature below which phase separation occurs and $p_c$ corresponds to the inflection point of the EOS at $T_c$ \cite{Yuan2006}. Substituting $p$ from Eq \ref{cs_eos} in Eq \ref{psi} and inserting the result into Eq \ref{fsc} cancels out $G$, leaving $T$ as the only parameter that controls the liquid to gas density ratio. Note that while the magnitude of $G$ becomes irrelevant, its sign is still of importance, therefore, we set $G=-1$. In the original LBM, the average fluid density, $\rho_0$, is directly set to unity, however, in a multiphase simulation using the modified $\psi(\rho)$, the fluid density is controlled by the EOS. We use $a=1$ lu, $b=4$ lu, and $R=1$ lu following Yuan and Schaefer \cite{Yuan2006} and $T=0.7Tc$, which results in a gas density, $\rho_g$, of $0.006$ lu and liquid density, $\rho_l$, of about $0.359$ lu. The C-S EOS with the selected parameters is shown in Figure \ref{cs_eos_fig}.

\subsection{Solid boundaries}\label{subsec3}

In the presence of solids inside the fluid domain, the solids are mapped onto the lattice. The solid nodes are treated as bounce-back boundary conditions \cite{Ginzbourg1994}, meaning that during the streaming stage, the updated distribution function pointing from a fluid node to a solid node is flipped in the opposite velocity direction rather than being streamed to the solid node. This results in a no-slip velocity condition at the fluid-solid boundary \cite{Kruger2017}.

The interaction force, $\boldsymbol{F}_{SC}$, for fluid nodes adjacent to solid nodes is similarly calculated using Eq \ref{fsc}, the only implication being that $\psi(\boldsymbol{x}+\boldsymbol{c}_i\Delta t)$ is based on the solid density, $\rho_s$. The magnitude of $\rho_s$ controls the contact angle between the solids and the liquid phase. $\rho_s = \rho_g$ creates a hydrophobic surface (contact angle of 180$^{\circ}$), while $\rho_s = \rho_l$ creates a hydrophilic surface (contact angle of 0$^{\circ}$) \cite{Kruger2017}. We set $\rho_s = 0.25$, which corresponds to a contact angle of 45$^{\circ}$ as shown in Section \ref{sec3}.

\section{Origin of matric suction}\label{sec3}

\begin{figure}[!tbp]
\centering
\includegraphics[width=0.525\textwidth]{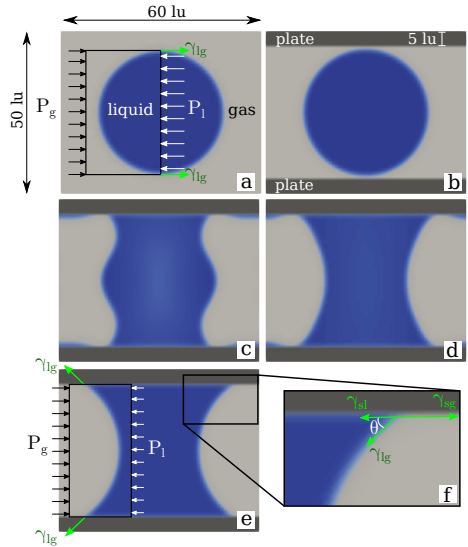}
\caption{2D simulation showing the origin of matric suction: a) liquid bubble at equilibrium inside a gas domain ($P_l>P_g$), b) placement of two solid plates on either side of the liquid bubble, c) liquid bubble being attracted to the solid, d) liquid bridge forming, e) liquid bridge at equilibrium ($P_g > P_l$), and f) equilibrium of surface tensions at the liquid-solid-gas boundary. Note that there is also an upward force with the magnitude $\gamma_{lg}sin(\theta)$, exerted by the solid to the liquid, that is not shown in subplot f, since it does not affect the contact angle.}
\label{suction_origin}
\end{figure}

We investigate how positive matric suction appears in a multiphase system by simulating the interaction between a liquid bubble and two parallel plates surrounding it, as shown in Figure \ref{suction_origin}. Initially, we place a liquid bubble (blue) inside a periodic gas domain (light gray) of size 60$\times$50 lu$^2$ and allow it to reach equilibrium in the absence of gravity (Figure \ref{suction_origin}a). At equilibrium, the liquid-gas surface tension ($\gamma_{lg}$) on the outer surface of the liquid bubble causes the liquid pressure ($P_l$) inside the bubble to be greater than the outer gas pressure ($P_g$). The pressure difference ($\Delta P$) follows the Young-Laplace equation,
\begin{equation}
\Delta P = \gamma_{lg}(\frac{1}{R_1}+\frac{1}{R_2}),
\label{laplace}
\end{equation}
where $R_1$ and $R_2$ are the principal radii of curvature of the liquid-gas interface. For this case, $R_1$ is the radius of the liquid bubble (blue circle) and $R_2$ (the out-of-plane radius of curvature) is infinity because the bubble is a 2D circle. We then place two solid plates (dark gray), with thicknesses of 5 lu, at the top and bottom side of the liquid bubble (Figure \ref{suction_origin}b). As time progresses, the liquid bubble is attracted to the solid plates (Figure \ref{suction_origin}c) and gradually forms a liquid bridge (Figure \ref{suction_origin}d) until a new equilibrium state is reached (Figure \ref{suction_origin}e). For this new equilibrium state, two conditions need to be satisfied: (a) The boundary between the liquid, solid and gas (Figure \ref{suction_origin}f) needs to be at equilibrium in terms of solid-gas, solid-liquid, and liquid-gas surface tensions ($\gamma_{sg}$, $\gamma_{sl}$, and $\gamma_{lg}$). 

In this simulation, the properties of the solid are set such that it creates a hydrophilic surface ($\gamma_{sg}>\gamma_{sl}$). Therefore the corner of the liquid will keep extending outwards (Figure \ref{suction_origin} from c to d to e) and the contact angle, $\theta$, will keep decreasing until $cos(\theta)$ is large enough for $\gamma_{sg}-\gamma_{sl}=\gamma_{lg}cos(\theta)$ to be satisfied (Figure \ref{suction_origin}f). (b) The pressure values inside and outside the liquid bridge (Figure \ref{suction_origin}e) should be at equilibrium considering the surface tension (green arrows in Figure \ref{suction_origin}e). Since the liquid now has an outward curvature because of condition (a), the pressure inside the liquid ($P_l$) has to be lower than the outer gas pressure ($P_g$) for force balance in the horizontal direction (Young-Laplace equation). The gas pressure is almost constant throughout this simulation (similar to physical tests where the gas pressure is the atmospheric pressure), therefore, the liquid pressure has to drop, and that is achieved by a slight drop of the liquid density (see the EOS in Figure \ref{cs_eos_fig}). Given that the system is closed and the total fluid mass is conserved, the liquid density, and hence pressure, must decrease as a result of the liquid phase expanding slightly and occupying some of the gas phase during the formation of the liquid bridge (constant mass $\to$ larger liquid volume $\to$ lower liquid density $\to$ lower liquid pressure). 

The entire mechanism can be thought of as the surface tension pulling against the corners of the liquid and extending it until: (a) the contact angle satisfies the equilibrium of surface tensions given by $\gamma_{sg}-\gamma_{sl}=\gamma_{lg}cos(\theta)$ and (b) the liquid density, hence pressure, is low enough to satisfy the mechanical equilibrium as per the Young-Laplace equation (Eq \ref{laplace}). Therefore, in the presence of hydrophilic solids, the liquid pressure drops below the gas pressure and positive matric suction ($\Delta P = P_g - P_l$) appears. 

For the 2D bridge-between-plates model shown in Figure \ref{suction_origin}, the magnitude of the matric suction is independent of the amount of liquid in the system, as long as the liquid remains as a bridge. This point can be deduced from the Young-Laplace equation (Eq \ref{laplace}, where the magnitude of the matric suction is a function of the liquid-gas surface tension and the mean radius of curvature. Given that (a) the surface tension is a constant material property, (b) the mean radius of curvature for a meniscus between 2 plates in 2D is only a function of the distance between the plates (2R) and the contact angle ($R_1=R/cos(\theta)$ and $R_2=\infty$), and (c) the contact angle at equilibrium is also a constant material property, the suction must only be a function of the distance between the two plates. We demonstrate this point by changing the amount of liquid in the system from the equilibrium state of the bridge in Figure \ref{suction_origin}e, both by injecting liquid into the bridge and by draining liquid out of the bridge, and measuring the matric suction. We perform injection/drainage of liquid by increasing/decreasing the liquid density by a very small amount and allowing the fluid to redistribute and reach a new equilibrium state. Once the system reaches the new equilibrium, we calculate the matric suction as the difference between the bulk liquid pressure and the bulk gas pressure, where bulk refers to the zone away from the liquid-gas boundary. 

\begin{figure}[t]
\centering
\includegraphics[width=0.525\textwidth]{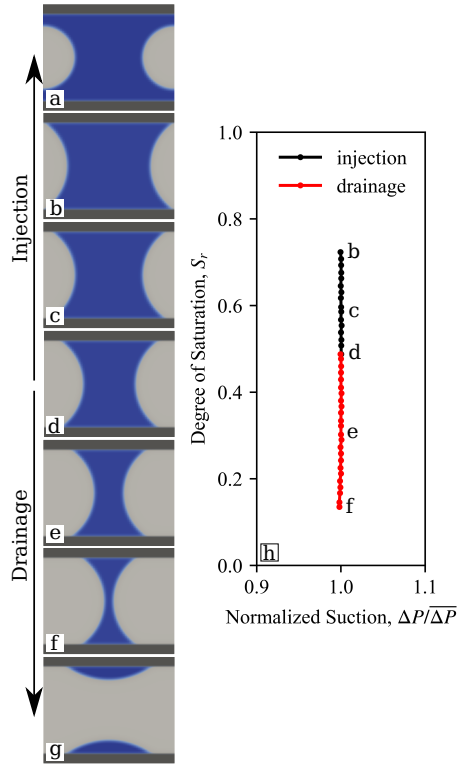}
\caption{2D simulation showing that suction is constant for a liquid bridge between two plates. The simulation is initialized from Point d which is equivalent to Figure \ref{suction_origin}e, and the degree of saturation, $S_r$, is changed by injection (Points c to a) and drainage (Points e to g). The suction and $S_r$ values for these points are labeled in panel h. The suction is normalized by the average of all suctions from Point b to f. The suction for Points a and g are not shown since the liquid is not in the form of a bridge.}
\label{constant_suction}
\end{figure}

In Figure \ref{constant_suction}, the measured (matric) suction ($\Delta P$) normalized by the average of all measured (matric) suctions from Point b to f ($\overline{\Delta P}$) is shown as a function of degree of saturation ($S_r$ = liquid volume / total pore volume). We observe that the suction magnitude remains constant between $S_r$ of about 15$\%$ to 75$\%$ where the liquid is in the form of a bridge between the two plates (Figure \ref{constant_suction}b to \ref{constant_suction}f). At $S_r$ above 75$\%$ (Figure \ref{constant_suction}a) the menisci lose contact with the plates and form gas bubbles, and at $S_r$ below 15$\%$ (Figure \ref{constant_suction}g) the bridge splits into two liquid droplets; these are not regions of interest in this study and hence their corresponding suctions are not shown in Figure \ref{constant_suction}. In addition, Figures \ref{constant_suction}b to \ref{constant_suction}f show that the equilibrium contact angle is constant regardless of $S_r$ level and the drainage/injection path. 

\begin{figure}[t]
\centering
\includegraphics[width=0.525\textwidth]{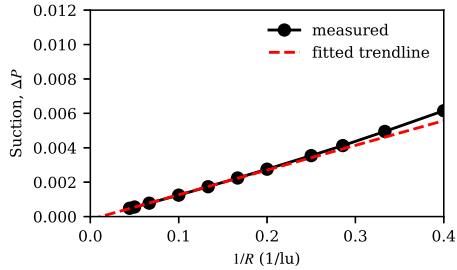}
\caption{The Young-Laplace test for finding the numerical surface tension using the bridge-between-plates model. $R$ is half the distance between the plates. The slope of the trendline fitted is $\gamma_{lg}cos(\theta)$.}
\label{lapace_test}
\end{figure}

We can also use the 2D bridge-between-plates model to measure the surface tension in our numerical simulation, by varying the spacing between the plates to create menisci with different curvatures and measuring the corresponding matric suction. Figure \ref{lapace_test} shows matric suction as a function of $1/R$, where $R$ is half the distance between the plates. Each point on the black curve has been acquired from a simulation similar to the one shown in Figure \ref{suction_origin}, but with a different plate spacing. Based on Eq \ref{laplace} and the fact that for the 2D case, $R_1=R/cos(\theta)$ and $R_2=\infty$, we can deduce that the slope of the trendline fitted to the curve in Figure \ref{lapace_test} is $\gamma_{lg}cos(\theta)$, measured as 0.0144 lu for our simulations. We can also measure the equilibrium contact angle graphically, from visualizations similar to Figures 1e and \ref{suction_origin}f, as 45$^\circ$, therefore resulting in $\gamma_{lg}$ of 0.0204 lu. Note that the measured curve deviates from the fitted line as $1/R$ increases (i.e. distance between the plates decreases); particularly, at $1/R$ of above 0.3 (distance of 7 grid points) the deviation becomes more obvious, showing the effect of grid resolution on the multiphase behavior. 

As mentioned in Section \ref{subsec1}, if comparing the simulation results to a physical test is of interest, the simulation lattice units can be converted to physical units. However, in this study, as only the qualitative behavior of the multiphase system is of interest, we do not convert the surface tension and suction to physical units. Instead we use the value found for $\gamma_{lg}$ to normalize the suction measured in lattice units. This normalized suction corresponds to the mean menisci curvature (inverse of mean radius of curvature) in units of 1/lu.

\section{Change of matric suction with degree of saturation}\label{sec4}

\begin{figure}[t]
\centering
\includegraphics[width=0.525\textwidth]{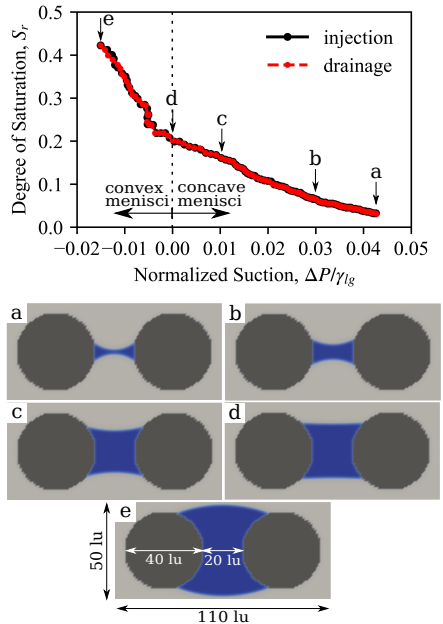}
\caption{2D simulation of a liquid bridge between two grains, showing the change of suction with $S_r$ and its insensitivity to injection/drainage path. From Point a to Point c the menisci are concave and the suction is positive, at Point d the menisci become planar with zero suction, and at Point e the menisci become convex resulting in a negative suction value.}
\label{two_disk}
\end{figure}

In the previous section, we show that the suction at equilibrium remains constant for a liquid bridge between solid plates, regardless of the amount of liquid in the system. However, that is not the case for liquid between non-planar surfaces, where the amount of liquid controls the mean radius of curvature of the meniscus and, in turn, the suction. We illustrate this point by now simulating a liquid bridge between two circular grains in 2D, shown in Figure \ref{two_disk}. We initially inject liquid into the bridge, from Point a to e, labeled in the suction-$S_r$ curve in \ref{two_disk}, and subsequently drain the liquid, from Point e to a, using a similar procedure explained in the previous section. We can see in Figure \ref{two_disk} that as $S_r$ increases, the normalized suction, which corresponds to the curvature of the menisci as per the Young-Laplace equation, decreases. The suction is positive while the menisci are concave (i.e. Points a, b, and c), reaches a value of zero when the menisci become planar (i.e. Point d) and switches to negative values when the menisci become convex (i.e. Point e). The small oscillations in the suction-$S_r$ curve are due to the non-smooth surface of the grains as a result of step-wise discretization. In addition to the change of suction with $S_r$, this simulation shows that for the case of a liquid bridge between two grains, the suction-$S_r$ relationship is independent of the injection/drainage path (i.e., there is no hysteresis).

\section{Origin of hysteresis in the suction-saturation relationship}\label{sec5}

In the simple two grains model shown above, we saw that the magnitude of suction at a given $S_r$ is independent of drainage or injection paths. Nevertheless, path-dependence or hysteresis does in fact occur for granular systems comprising many grains \cite{Pham2003, Likos2014}. To understand the origin of hysteresis, we will first simulate drainage and injection experiments for a 2D granular packing with only 15 grains, where we can easily visualize the liquid and gas cluster formations and meniscus curvatures. Once we have established the basic concepts in 2D, we will simulate drainage and injection experiments for a 3D granular packing with 1068 grains and apply the same concepts to 3D.
\subsection{2D granular packing}\label{2d_granular}

\begin{figure}[t]
\centering
\includegraphics[width=0.438\textwidth]{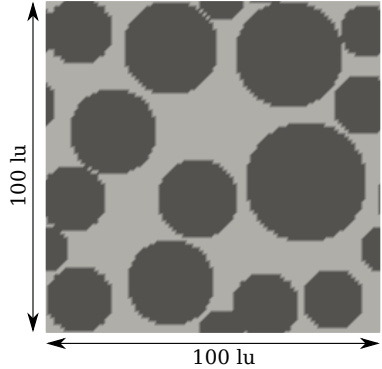}
\caption{The 2D granular packing model, consisting of 15 grains with an average diameter of 22 lu. The boundaries are periodic in both directions.}
\label{2d_packing}
\end{figure}

Consider the 2D granular packing, shown in Figure \ref{2d_packing}, formed by 15 circular grains, with an average diameter of 22 lu, inside a 100$\times$100 lu$^2$ domain with periodic boundaries in both directions. The pore structure consists of a number of chambers (wider pore spaces) connected through throats (narrower pore spaces). The grains are fixed in position and only the fluid domain is simulated. Gravity is not applied. The simulations in this section consist of a full cycle of drainage and injection. We initialize the simulation with a very small gas bubble in the pore space, corresponding to an $S_r$ of 97$\%$. Subsequently, we slowly drain the liquid out of the system down to an $S_r$ of 1$\%$. From this almost dry state, we inject the liquid back into the system until full saturation. 

We use a drainage/injection procedure similar to what was used in the previous section; we drain/inject the liquid by reducing/increasing the liquid density everywhere in the system by a small amount and allowing the fluid to redistribute until it reaches a new equilibrium state. We adjust the density decrement/increment such that approximately the same volume of liquid is drained/injected at each drainage/injection step, corresponding to an almost linear change of $S_r$. Only at $S_r<0.1$, to avoid instability, we switch to using constant density decrement/increment, which corresponds to an exponential decrease of $S_r$. 

\begin{figure}[t]
\centering
\includegraphics[width=0.525\textwidth]{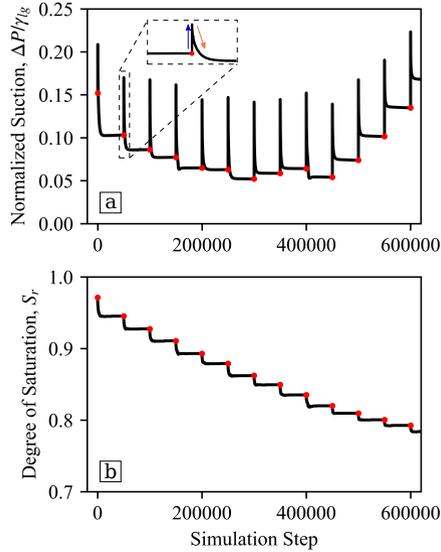}
\caption{a) Suction and b) $S_r$ as a function of simulation step (time) for the first 600,000 steps of the drainage simulation of the 2D granular packing model. The inset of subplot (a) shows a zoomed-in view of a drainage step. The blue arrow shows the sudden increase of suction upon drainage, while the orange arrow shows the more gradual decrease of suction towards equilibrium. The red circles correspond to equilibrium points.}
\label{drain_step}
\end{figure}

Figure \ref{drain_step} shows the change of suction and $S_r$ with time for the first 600,000 steps of the drainage simulation. When a density decrement is applied, the liquid pressure abruptly drops, causing a sudden increase of suction (blue arrow in Figure \ref{drain_step}a inset). This sudden change of suction creates a non-equilibrium condition for the bubbles or menisci, where there is an unbalanced force pushing outwards from the gas zone towards the liquid zone. As a result the gas zone starts expanding, causing a decrease in liquid volume, increase in liquid density and liquid pressure, and decrease in suction (orange arrow in Figure \ref{drain_step}a inset). The value at which the suction stops decreasing and reaches equilibrium again depends on the pore structure and the possible curvatures the bubble or menisci can take. As seen in Figure \ref{drain_step}a, in some cases the suction reaches equilibrium at a lower value than the previous step, while in other cases it reaches equilibrium at a higher value. Although not shown here, the same mechanism is true for injection, only in the opposite direction. With the injection of liquid, the liquid pressure abruptly increases, causing a sudden decrease of suction, and creating an unbalanced outward force from the liquid zone towards the gas zone. The unbalanced force causes the liquid zone to expand, during which the liquid pressure decreases and the suction increases. Again, the value at which the suction reaches equilibrium depends on the pore structures and the possible curvatures the bubble or menisci can take. 

\begin{figure}[t]
\centering
\includegraphics[width=0.525\textwidth]{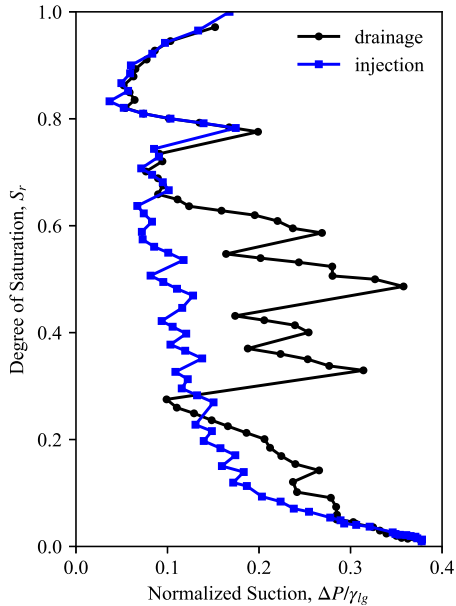}
\caption{Soil-water characteristic curves for drainage and injection simulations of the 2D granular packing model.}
\label{SWCC_2D}
\end{figure}

We use the suction and $S_r$ values at equilibrium points (the red circles in Figure  \ref{drain_step}) to form the suction-$S_r$ curve, i.e., soil-water characteristic curve (SWCC) for both drainage and injection simulations of the 2D granular packing model. The SWCCs are shown in Figure \ref{SWCC_2D}. Unlike the example of liquid bridge between two grains, where the suction does not vary between the injection and drainage paths, we see a hysteresis in the SWCC for the small 2D granular packing. The suction at a given $S_r$ depends on the drainage/injection path. To understand the shape of each curve and, more importantly, the source of the hysteresis, we separately investigate the pore emptying and pore filling processes in detail, followed by a comparison of the two processes.

\subsubsection{The pore emptying process}\label{2d_emptying}

\begin{figure*}
\noindent
\begin{minipage}[t]{0.63\columnwidth}
    \includegraphics[width=\textwidth]{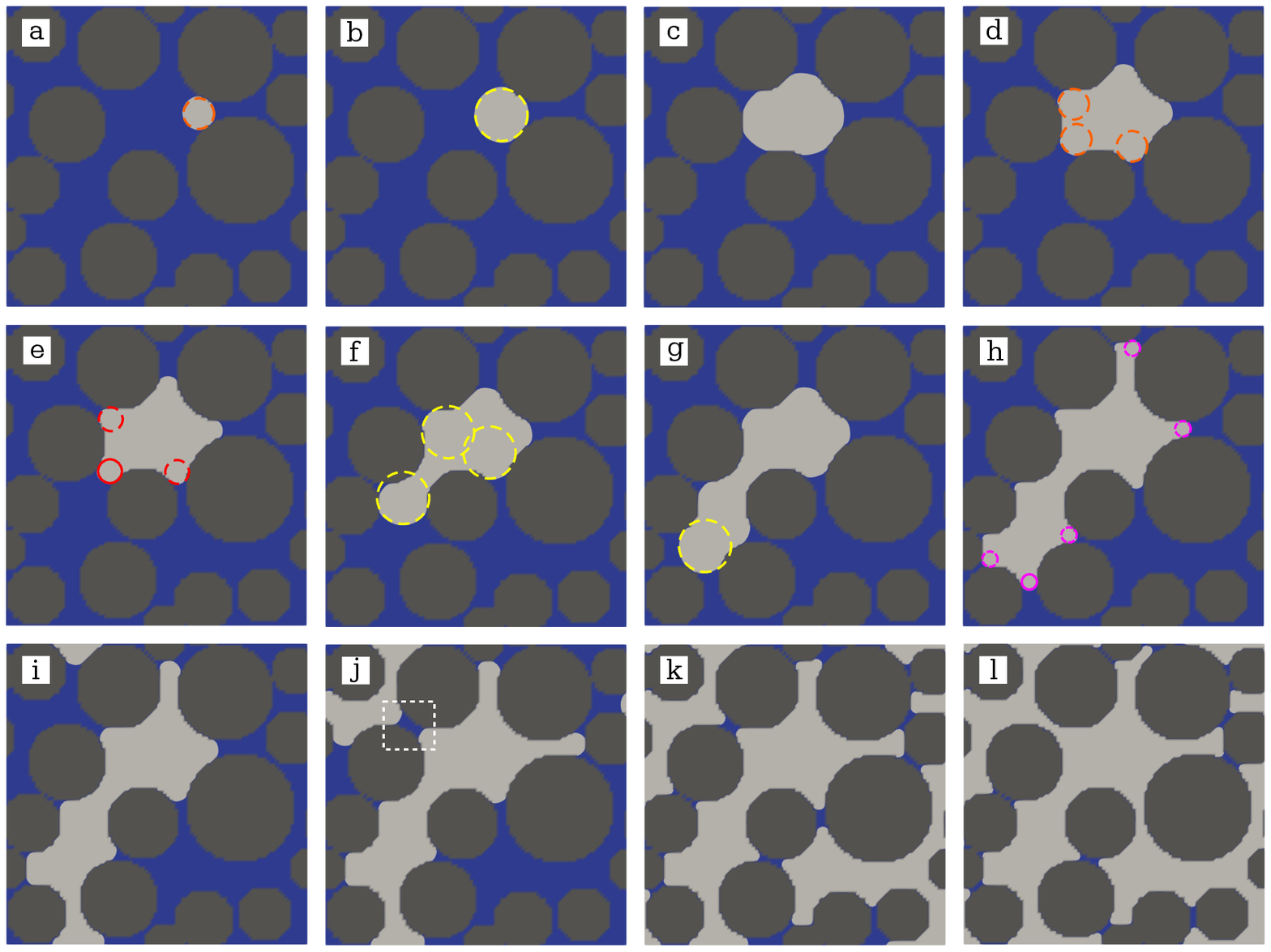}
    \caption{Snapshots of the drainage simulation for the 2D granular packing model at different equilibrium points during the pore emptying process. The gas and liquid phases have been binarized for easier visualization and are shown with light gray and blue, respectively. Unfilled circles have been fitted to some of the menisci to visually assist with comparing the radii of curvature between different snapshots. Circles of the same color have approximately the same radius. The pink-colored circles have the smallest radius followed by red, orange, and yellow. The corresponding suction to each snapshot is labeled in Figure \ref{drainage}. 
    }
    \label{pore_emptying}
\end{minipage}%
\hfill%
\begin{minipage}[t]{0.34\columnwidth}
    \includegraphics[width=0.97\textwidth]{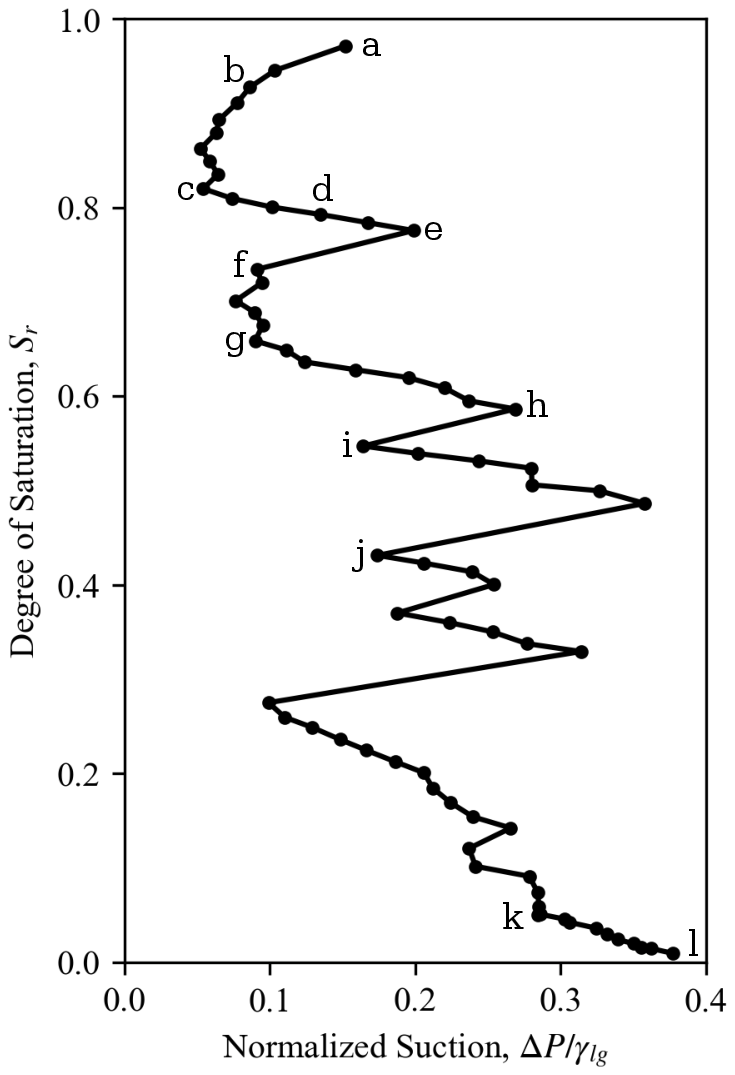}
    \caption{Soil-water characteristic curve for the drainage simulation of the 2D granular packing model with labeled points corresponding to the snapshots in Figure \ref{pore_emptying}.}
\label{drainage}
\end{minipage}
\end{figure*}

We use snapshots of the simulation at different equilibrium points during the drainage process and their corresponding suction to discuss the pore emptying process. Figure \ref{pore_emptying} includes these snapshots, labeled a through l, and the corresponding suction values are marked in Figure \ref{drainage}.  We have added circles to some of the menisci in Figure \ref{pore_emptying}, with radii consistent with the corresponding suction values, to visually assist with comparing the radii of curvature between different snapshots. 

With the initiation of drainage at Point a, the gas bubble starts expanding into the available pore space around it. As the bubble grows to Point b, the radius of curvature increases, and therefore the suction at equilibrium decreases. The expansion continues to Point c, where the gas bubble fully takes over the chamber where it is located. At this point the suction reaches a local minimum. Further expansion of the bubble requires the liquid menisci to retreat at the throats. As the menisci retreat into the throats (i.e. the gas zone advances into the throat) at Point d, the radius of curvature decreases (since the throats are narrowest at the center) and therefore the suction at equilibrium increases. Since the liquid zone is continuous at this point, the liquid pressure is evenly distributed at equilibrium and, as a result, all the menisci have the same radius of curvature. This is the case for all points above $S_r$ of 0.4 in this simulation. We have not fitted all menisci with a circle to allow better visualization. 

The increase of suction continues up to Point e, where the meniscus in the widest throat reaches the narrowest part of that throat (see bottom-left circle in Figure \ref{pore_emptying}e shown with a solid line). At this point, the suction has reached the air-entry value (AEV) of the neighboring chamber, and with further drainage the gas enters the neighboring chamber at Point f. Since the chamber now allows a wider meniscus to form, the suction drops considerably. Notice that the menisci at the other throats, which had previously retreated into the throats, now advance to create the larger radius of curvature required for equilibrium (compare Figure \ref{pore_emptying}e and \ref{pore_emptying}f). From Point f to Point g, the gas gradually takes over the new chamber, and the radius of curvature, hence suction, remains nearly constant. When the new chamber is completely emptied of liquid at Point g, further expansion of the gas zone would again require the liquid menisci to retreat at the throats, similar to the transition from Point c to d. 

At Point h, the meniscus in the widest throat reaches the narrowest part of that throat (see bottommost circle in Figure \ref{pore_emptying}h shown with a solid line) and the suction becomes the AEV of the neighboring chamber, resulting in the emptying of that chamber at Point i (Note that since the boundaries are periodic, the continuation of the chamber at the bottom appears at the top). Pore emptying continues with the same process repeating between Points i and k. Notice how at Point j, the emptying of a chamber results in the formation of an isolated bridge (see the box in Figure \ref{pore_emptying}j). At Point k, all the major chambers have been emptied and the liquid is only in the form of bridges between particles. At this point, further drainage would require the liquid bridges to shrink, resulting in smaller radii of curvature. Therefore, from this point forward, the suction strictly increases.

Based on the observations above, we can describe the change of matric suction during drainage using two pore emptying stages: 1) the emptying of the chambers and 2) the emptying of the liquid bridges at the throats. The first stage consists of a series of repeating patterns in terms of change of suction, where the suction drops considerably upon the entry of the gas phase into a chamber, remains rather constant or continues to decrease as the gas takes over the chamber, and finally increases again as the menisci retreat in the throats. At the end of the first stage, all major chambers have been emptied and the remaining liquid is in the form of liquid bridges between two grains. During the second stage, where the liquid bridges are removed by drainage, the suction continuously increases with the reduction of $S_r$.

An interesting observation is that chambers do not necessarily empty in the order of descending size or increasing AEV. Other than the chamber where the initial gas bubble resides, a chamber that empties has the following conditions: (a) it is adjacent to a chamber that has already been fully emptied and (b) the throat that connects it to the empty chamber is the widest among all the connecting throats of the other chambers that meet condition (a). This chamber is not necessarily the largest available. For instance, there could be a large chamber that has a very narrow connecting throat (large AEV), preventing it from emptying before other smaller chambers. On the other hand, there could be a chamber that has a very wide connecting throat (small AEV) but located far away from the expanding gas zone, therefore emptying later in the process.

\subsubsection{The pore filling process}\label{2d_filling}

\begin{figure*}
\noindent
\begin{minipage}[t]{0.63\columnwidth}
\includegraphics[width=1\textwidth]{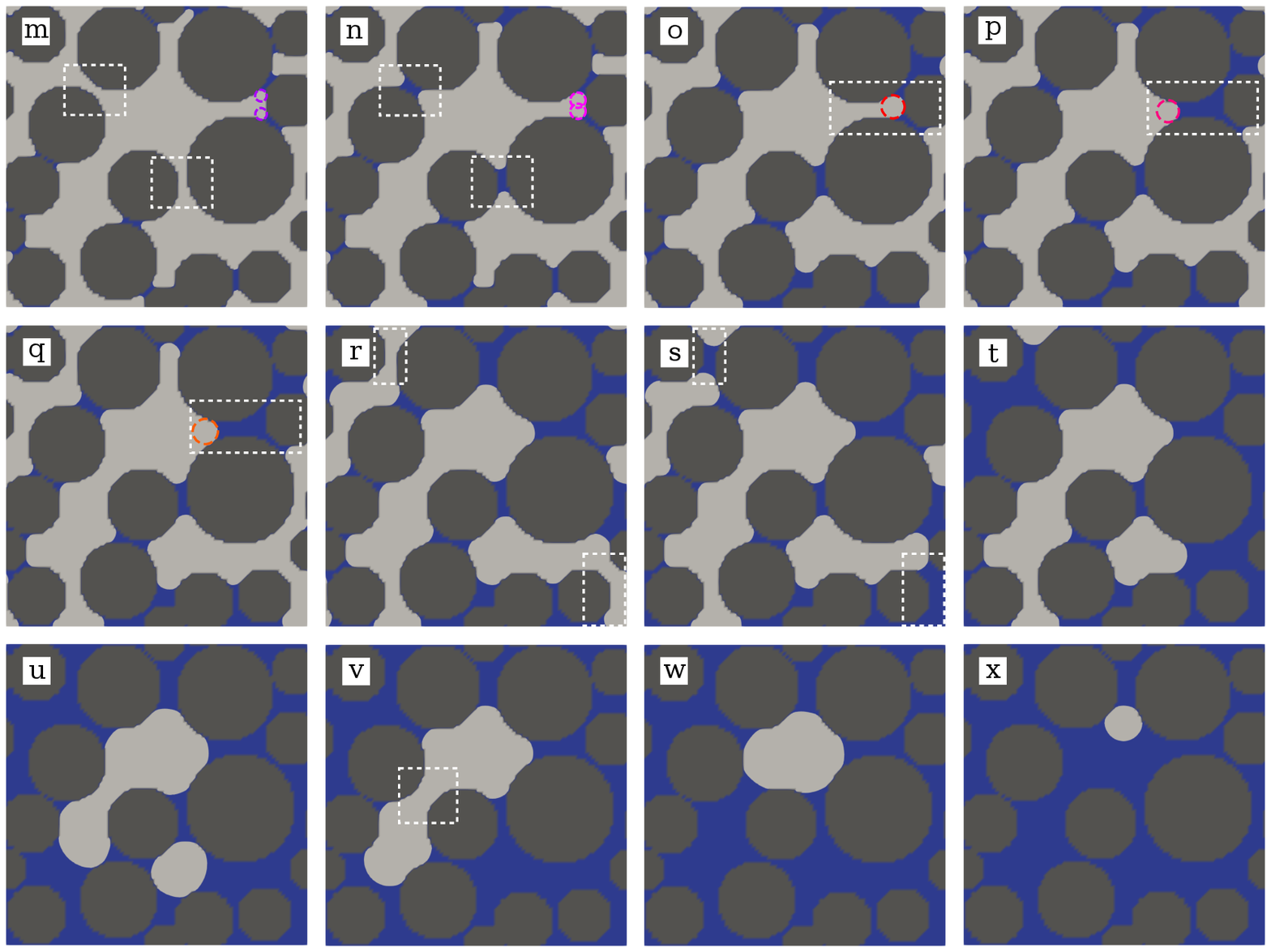}
\caption{Snapshots of the injection simulation for the 2D granular packing model at different equilibrium points during the pore filling process. The gas and liquid phases are shown with light gray and blue, respectively. Unfilled circles have been fitted to some of the menisci to visually assist with comparing the radii of curvature between different snapshots. Circles of the same color have approximately the same radius. The purple-colored circles have the smallest radius followed by pink, rose, red, and orange. The corresponding suction to each snapshot is labeled in Figure \ref{injection}. Refer to the text for the purpose of the white boxes.}
\label{pore_filling}
\end{minipage}%
\hfill%
\begin{minipage}[t]{0.34\columnwidth}
\includegraphics[width=0.97\textwidth]{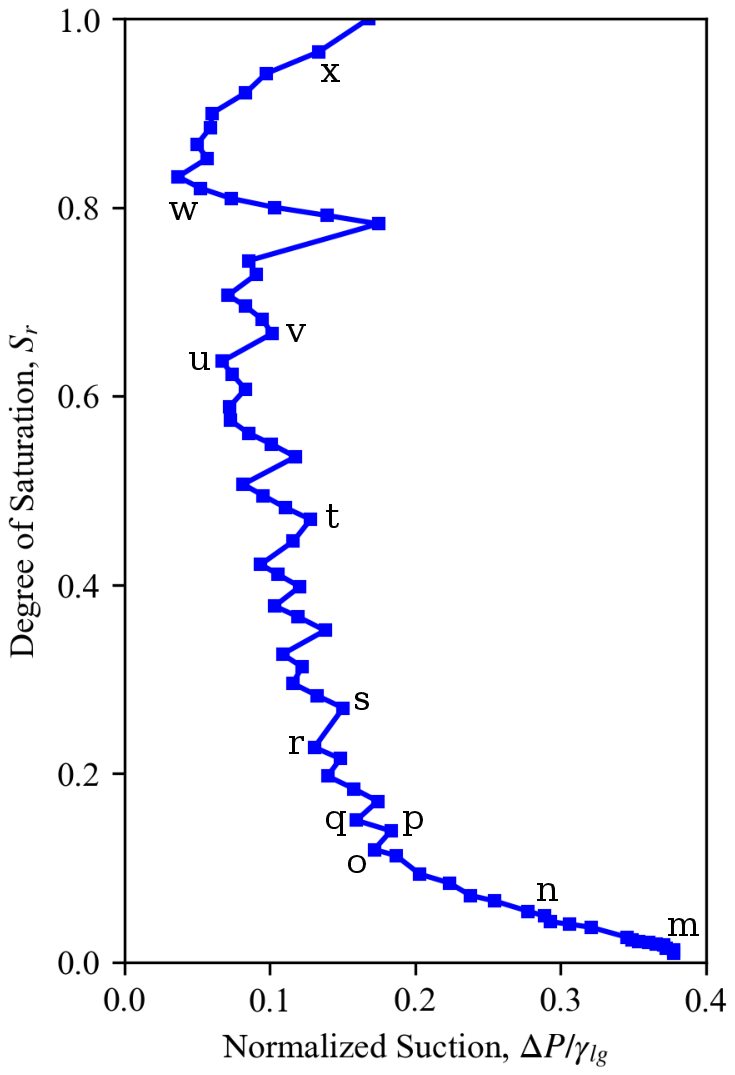}
\caption{Soil-water characteristic curve for the injection simulation of the 2D granular packing model with labeled points corresponding to the snapshots in Figure \ref{pore_filling}.}
\label{injection}
\end{minipage}%
\end{figure*}

Similar to the previous section, we use snapshots of the simulation, shown in Figure \ref{pore_filling}, and their corresponding suctions, marked in Figure \ref{injection}, to discuss the pore filling process. We have labeled the injection Points from m to x, to avoid confusion with the drainage points. 

We start the injection simulation from Point m which is the same as Point l of the drainage simulation. With the initiation of injection, the already existing liquid bridges start expanding, which results in an increase in radius of curvature and decrease in suction (see circles growing in size from Figure \ref{pore_filling}m to \ref{pore_filling}n), similar to what we saw with the two grains model in Figure \ref{two_disk}a to \ref{two_disk}c. In addition to the expansion of existing bridges, the gas phase starts condensing at empty throats, forming new liquid bridges (see boxes in Figures \ref{pore_filling}m and \ref{pore_filling}n). The same process continues up to Point o. From Points m to o, the expansion of liquid bridges results in a continuous decrease of suction. At Point o, the bridges have expanded enough for two bridges to join together at the smallest chamber in the system (see the box in Figure \ref{pore_filling}o). The meniscus radius of curvature that is formed in this chamber at Point o is the largest possible for this chamber, and any further injection and expansion of the liquid zone would require the meniscus to take a smaller radius of curvature, therefore, the suction slightly increases at Point p. As soon as the meniscus crosses the narrowest part of this chamber, it can take larger radii of curvature again, and so the suction drops for Point q. The same process keeps repeating as other bridges coalesce to form a meniscus which then fills a chamber. Another example of this process is shown at Points r to s, in the bottom right box. 

An additional event that occurs at Point s is the formation of a new liquid bridge, in the top left box, which creates three separate gas zones in the system (note that the domain is periodic). We refer to these separate gas zones as gas clusters. As the injection and liquid zone expansion continues, these gas clusters shrink simultaneously, and eventually the smallest one dissolves in the liquid at Point t, leaving only two gas clusters. Similarly, from Point u to v, the next smallest gas cluster collapses with the expansion of the liquid zone. In general, the collapse of a gas cluster results in a slight increase of suction due to the drop in liquid density. From Points v to x, where there is only one gas cluster remaining, the system behaves similar to steps a to h of the drainage process, only in the opposite direction.

Based on the observations above, we can describe the change of matric suction during injection using three pore filling stages: 1) filling of the throats, 2) filling of the intermediate chambers, and 3) filling of the largest chamber. During the first stage, the existing liquid bridges expand, while new liquid bridges form as a result of capillary condensation at the empty throats. At this stage the suction continuously decreases. During the second stage, the expanding bridges join together and fill the chambers. Since all bridges are expanding simultaneously, the ones that are surrounding a smaller chamber join together first to fill the chamber, resulting in chambers filling in order of increasing size. Therefore, at this stage, with the increase of $S_r$, the suction has a general decreasing trend with small oscillations. The small oscillations occur either when a meniscus pushes through a chamber and fills it, or when a gas bubble collapses inside a chamber. Note that the large oscillation of suction happening at $S_r$ of about 0.8 is due to the specific pore structure in our model; if the two grains forming the throat shown in the box in Figure \ref{pore_filling}v were slightly closer together, then gas would have condensed and formed a liquid bridge at that throat, creating two separate gas clusters, and the jump in suction we see at $S_r$ of about 0.8 would not have occurred. During the last stage, which is the filling of the largest chamber, the suction gradually increases as the gas cluster becomes smaller. Once the final gas cluster collapses, suction instantly drops to zero. The reason why there is a gradual increase of suction in the last stage that we do not see in the previous stage is that the last gas cluster shrinks gradually while the other gas clusters suddenly collapse. To understand why that is, let us consider Point u. For the smaller gas cluster to be stable with an even smaller radius, it would require a high suction, however, the larger gas cluster at Point u does not need such a high suction. Therefore, the smaller gas cluster becomes unstable, quickly shrinks and collapses, while the larger gas cluster slightly expands. In other words, when more than one gas cluster is available, the larger clusters act as stabilizers for suction, while when there is only one gas cluster left, there is no such stabilization effect and the suction increases as the gas cluster shrinks.

Unlike the pore emptying process, where the conditions for a chamber to empty is independent of the chamber size, in the pore filling process we see that the chambers do actually fill in order of increasing size. Also, if we define the smallest suction value that a gas cluster in a chamber can hold before collapsing as the air-expulsion value (AEV) of the chamber (for instance the suction values at Points o or r),  we can conclude that the chambers do fill according to their AEV.

\subsubsection{Comparison of pore emptying and pore filling processes: source of hysteresis}\label{2d_comparison}

\begin{figure}[t]
\centering
\includegraphics[width=0.475\textwidth]{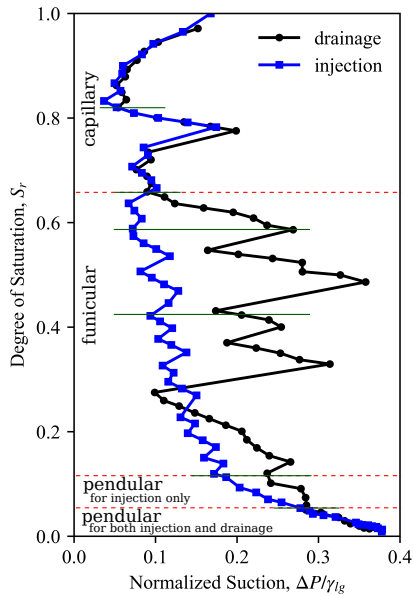}
\caption{Soil-water characteristic curves for drainage and injection simulations of the 2D granular packing model with the addition of liquid content states indicators. The red dashed lines separate the different liquid content states. The solid green lines correspond to the $S_r$ levels shown in Figure \ref{comparison_2d}.}
\label{states}
\end{figure}

The behavior of the wet granular material is usually described using three main liquid content states \cite{Iveson2001,Mitarai2006}: the pendular state, where the liquid is in the form of bridges between grains, the funicular state, where liquid-filled pores and liquid bridges coexist, and the capillary state, where all grains are immersed in liquid but suction still exists in the system due to the presence of gas bubbles. If we partition the SWCCs based on the definition of these states, as shown in Figure \ref{states}, we see that the drainage and injection SWCCs are almost identical in the mutual pendular and capillary states, and hysteresis only occurs in the funicular state.

\begin{figure}[!h]
\centering
\includegraphics[width=0.57\textwidth]{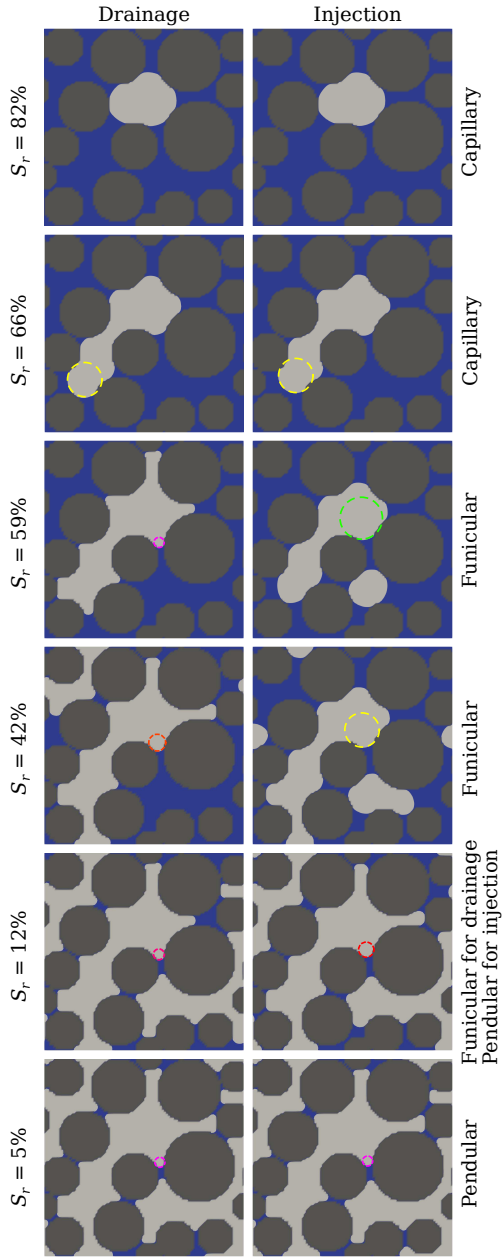}
\caption{Comparison of the gas and liquid distributions between drainage and injection simulations of the 2D granular packing model, at different $S_r$ levels.}
\label{comparison_2d}
\end{figure}

By comparing the pore emptying process with the pore filling process, we can find the source of hysteresis in the SWCC. In Figure \ref{comparison_2d}, we have plotted snapshots of the model at different $S_r$ levels for both drainage and injection simulations. For reference, the $S_r$ levels are marked with green solid lines in Figure \ref{states}. In the capillary state ($100\% > S_r > 66\%$) and the mutual pendular state ($S_r < 5\%$), the liquid and gas distributions, hence suction, are identical for drainage and injection paths (i.e., no hysteresis is observed). However, for the intermediate funicular state ($66\% > S_r > 5\%$), Figure \ref{comparison_2d} shows a larger radius of curvature for injection compared to drainage, at each $S_r$ level. This large difference in meniscus radius of curvature, and therefore suction, despite similar saturation level, is because during drainage there is only one gas cluster in the system at all times, while, during injection, multiple gas clusters are present. 

\subsection{3D granular packing}\label{3d_packing}

Based on the observations above, we can summarize our findings as follows.  When there is only one shrinking/expanding gas cluster in the largest chamber, i.e., capillary state, or there are only shrinking/expanding liquid bridges in the system, i.e. pendular state, the pore emptying and pore filling processes are similar, and therefore, there is no hysteresis. However, during the funicular state, where a series of chambers need to be filled or emptied, the pore emptying and pore filling processes are different. In the pore emptying process, the gas can only expand from the already existing gas zone and no new gas zones can appear inside the liquid zone, therefore, the gas zone has to push through the throats to empty the chambers, which requires a high suction. In the pore filling process, the gas can condense at throats and form new liquid zones inside the gas zone, creating many liquid bridges that expand simultaneously and coalesce to fill the chambers, therefore, there is no need to push through the throats in order to fill the chambers. The difference between the pore emptying and pore filling processes in the funicular state results in larger suction values and larger oscillations for drainage compared to injection.

\begin{figure*}[tbh]
\centering
\includegraphics[width=1\textwidth]{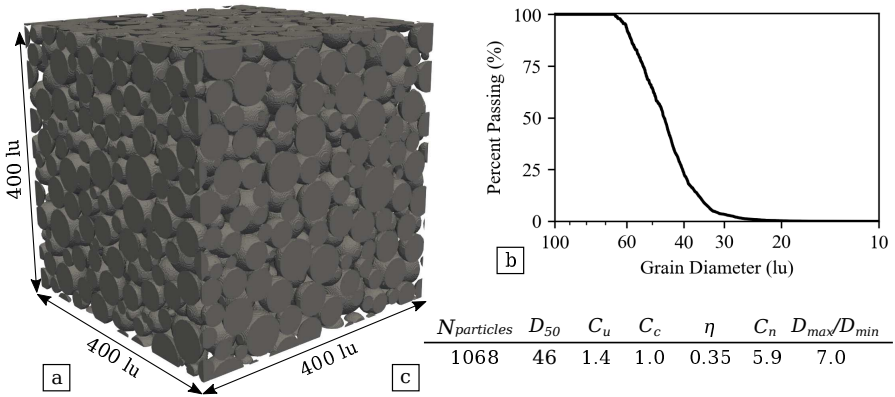}
\caption{3D granular packing model. a) Visualization of the granular packing. b) Grain size distribution curve. c) Statistics of the granular packing including total number of particles $N_{particles}$, diameter corresponding to $50\%$ passing on the grain size distribution $D_{50}$,  coefficient of uniformity $C_u = D_{60}/D_{10}$, coefficient of curvature $C_c = D_{30}^2/(D_{60}D_{10})$, porosity $\eta$ = total pore volume$/$total volume, coordination number $C_n = 2 \times$number of contacts$/N_{particles}$, and polydispersity $= D_{max}/D_{min}$.
}
\label{3d_config}
\end{figure*}

We utilize the Discrete Element Method (DEM) with PFC3D to create a stable 3D granular packing shown in Figure \ref{3d_config}a. We first create a fixed-size domain with periodic boundaries in all directions. Afterwards, we generate spherical grains, with sizes randomly drawn from the grain size distribution shown in Figure \ref{3d_config}b, and position them randomly throughout the specimen domain, until a target porosity ($\eta$ = total pore volume / total volume) is reached \cite{Itasca}. We then run a DEM simulation to ensure the grains are at equilibrium. We use the final equilibrium grain configuration from DEM, the statistics of which are presented in Figure \ref{3d_config}c, for the multiphase LBM simulation. We do not apply gravity in either the DEM or LBM simulations. Similar to the previous section, in the multiphase LBM simulation, grains are stationary and grain-grain interactions are not considered, implying the assumption that the induced suction and surface tension are not large enough to cause grain movement. Fixing the position of the grains allows us to eliminate the effects grain movement can have on the SWCC and focus on the source of hysteresis at constant porosity. 

\begin{figure}[tbh]
\centering
\includegraphics[width=0.475\textwidth]{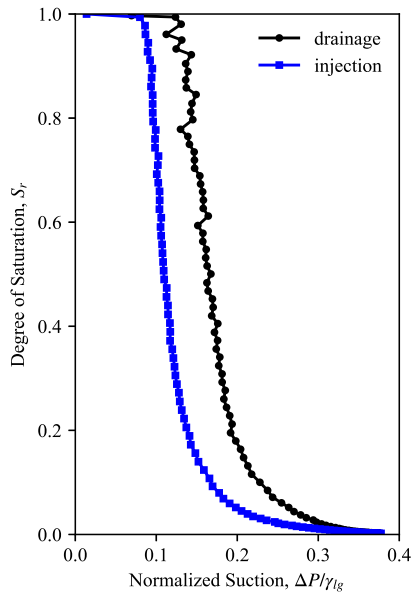}
\caption{Soil-water characteristic curves for drainage and injection simulations of the 3D granular packing model.}
\label{SWCC_3D}
\end{figure}

The drainage/injection procedure for the 3D model is identical to the 2D model explained in the previous section. The simulated SWCCs for a full cycle of drainage followed by injection are shown in Figure \ref{SWCC_3D}. We observe a clear hysteresis in the SWCC for the 3D model. Similar to the previous section, we compare the pore emptying with the pore filling process to identify the source of hysteresis. While for the relatively-simple 2D model we were able to visually follow and investigate the distribution of gas and liquid phases during pore emptying and pore filling, visualization is difficult for a large 3D model. Instead, we use a number of statistics describing the gas and liquid phase distributions to look into the pore emptying and pore filling processes. We identify the gas and liquid clusters, where a cluster refers to a disconnected zone of the same phase, at each step of the simulation, using the depth-first-search algorithm \cite{TARJANR1972}. We label each liquid cluster and use that information to find its order ($n$), defined as the number of grains connected to that liquid cluster \cite{Delenne2015}. We also track the volume of the largest liquid cluster in the system. Below, we first look at the pore emptying and pore filling processes separately, followed by a comparison between the two processes.

\subsubsection{The pore emptying process}\label{3d_pore_emptying}

\begin{figure*}[tbh]
\centering
\includegraphics[width=1\textwidth]{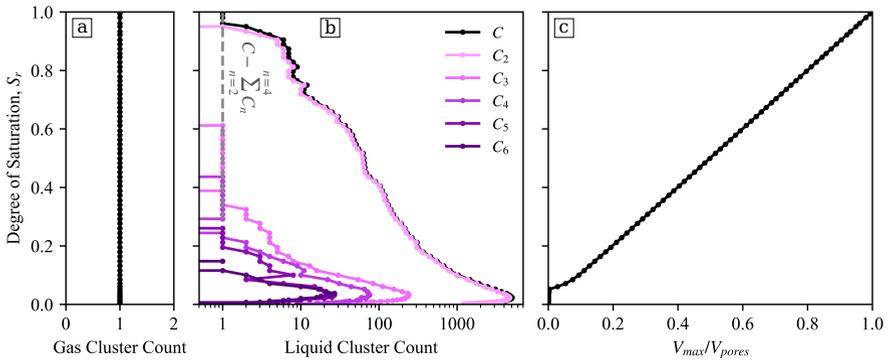}
\caption{Gas and liquid phase distribution statistics for the drainage simulation of the 3D granular packing model. Cluster refers to a disconnected zone of the same phase. $C$ includes all clusters, while $C_n$ only includes clusters with order $n$, which are clusters connected to $n$ grains. $V_{max}$ is the volume of the largest liquid cluster in the system, and $V_{pores}$  is the total volume of the pores.
}
\label{clusters_drain}
\end{figure*}

We use the gas and liquid cluster statistics plotted in Figure \ref{clusters_drain} to describe the pore emptying process. In Figure \ref{clusters_drain}a, we plot the gas cluster count and we see that there is only a single gas cluster during the entire pore emptying process, similar to the 2D model. In Figure \ref{clusters_drain}b, we plot the total liquid cluster counts ($C$), as well as the count of liquid clusters with different orders ($C_n$). We see that at the beginning of drainage there is only one liquid cluster in the system, but as the drainage progresses, the liquid cluster count increases rapidly. The majority of these clusters are of order two ($C_2$), which are liquid bridges between two grains. Liquid clusters of order three ($C_3$) and four ($C_4$) appear at $S_r$ below 60$\%$ and 40$\%$, respectively, and clusters of higher order appear as $S_r$ decreases further. If we subtract $C_2$, $C_3$ and $C_4$ from $C$, we see that there is one other liquid cluster contributing to the total number of liquid clusters. Our analysis shows that this cluster has an order between 1068, which is the total number of grains, and 1063, for the range plotted in Figure \ref{clusters_drain}b, meaning that this single liquid cluster is in contact with most of the grains in the system. In Figure \ref{clusters_drain}c we plot the volume of the largest liquid cluster, $V_{max}$, normalized by the total volume of pores, $V_{pores}$. We see that, for $S_r$ above 7$\%$,  $V_{max}/V_{pores}$ is almost equal to $S_r$, meaning that the single large cluster is what is making up the majority of the liquid volume and the clusters of order two, although high in count, have minimal contribution to the total volume of the liquid.

\begin{figure*}[tbh]
\centering
\includegraphics[width=1\textwidth]{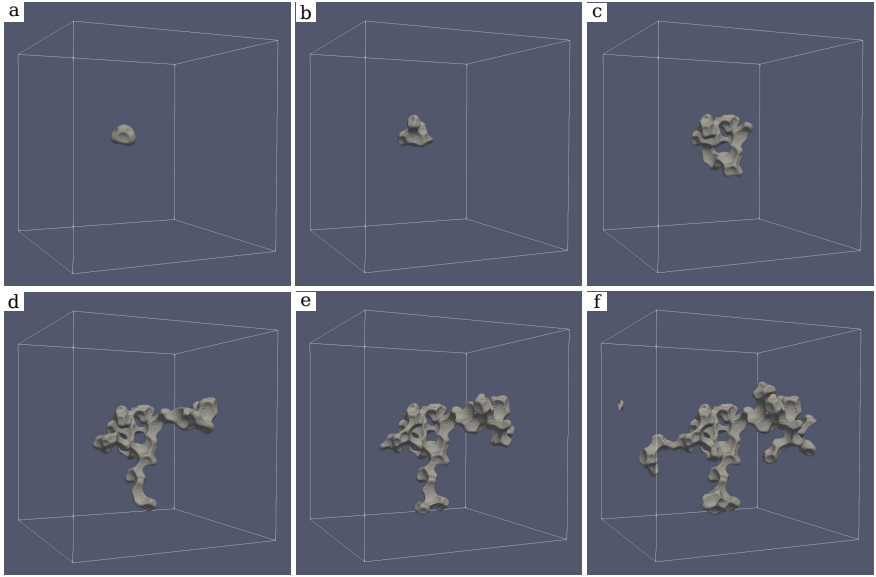}
\caption{The gas phase distribution for the first six drainage steps of the 3D granular packing model. The suction and $S_r$ values of these points are labeled in Figure \ref{increment_comparison}.}
\label{gas_cluster}
\end{figure*}

\begin{figure*}[!h]
\centering
\includegraphics[width=1\textwidth]{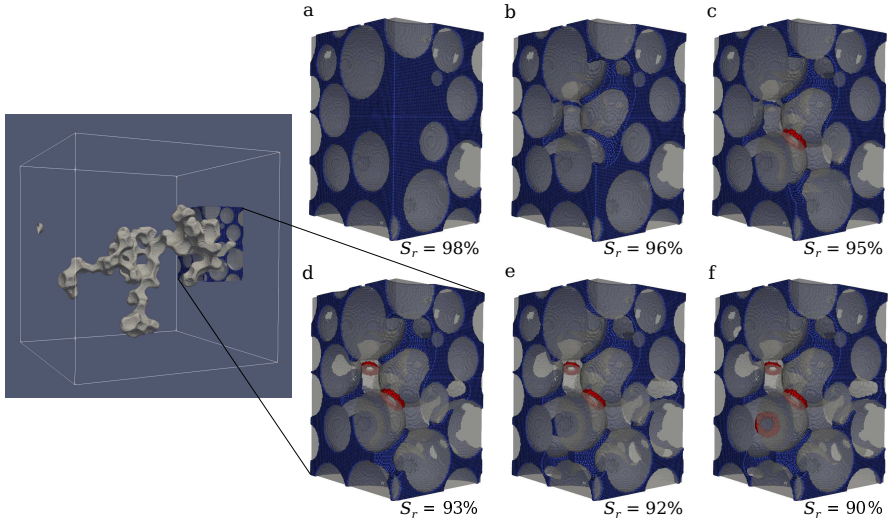}
\caption{The liquid phase distribution in a subdomain of the 3D granular packing model, at different $S_r$ levels during the drainage simulation. The single large liquid cluster is shown in blue, and the liquid bridges are shown in red.}
\label{liq_cluster}
\end{figure*}

Based on the observations above, we can summarize the pore emptying process as follows. In terms of the gas phase distribution, there is always a single gas cluster, expanding and pushing into adjacent throats in order to empty the adjacent chambers. Figure \ref{gas_cluster} shows the evolution of this gas cluster for the first six drainage steps. The process is slightly more detailed in terms of the liquid phase distribution. At $S_r$ above 96$\%$, there is only one large liquid cluster, immersing all the grains. At $S_r$ between 96$\%$ and 8$\%$, there is still one large liquid cluster that makes up the bulk of the liquid, but there are also an increasing number of smaller liquid clusters, most of which are liquid bridges. See for instance the subdomain shown in Figure \ref{liq_cluster} where the single large liquid cluster is shown in blue and the liquid bridges are shown in red. At $S_r$ between 8$\%$ and 5$\%$ the single large cluster splits into smaller clusters, again most of which are liquid bridges. At $S_r$ between 5$\%$ and 2$\%$ the number of liquid bridges continues increasing and finally, at $S_r$ below 2$\%$, the liquid bridges are drained. The pore emptying process we see here is consistent with what we had previously identified in 2D.

\begin{figure}[tbh]
\centering
\includegraphics[width=0.475\textwidth]{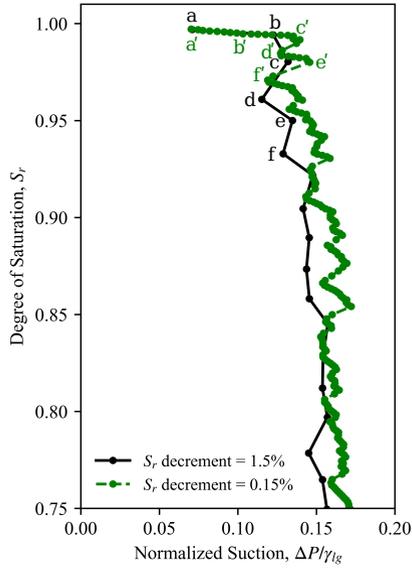}
\caption{Comparison of the drainage soil-water characteristic curves with different $S_r$ decrements (rate of drainage). The solid black curve is the same drainage curve shown in Figure \ref{SWCC_3D}. The points labeled on the solid black curve correspond to the subplots in Figure \ref{gas_cluster}, while the points labeled on the dashed green curve correspond to subplots in Figure \ref{gas_cluster_2}.}
\label{increment_comparison}
\end{figure}

\begin{figure*}[!h]
\centering
\includegraphics[width=1\textwidth]{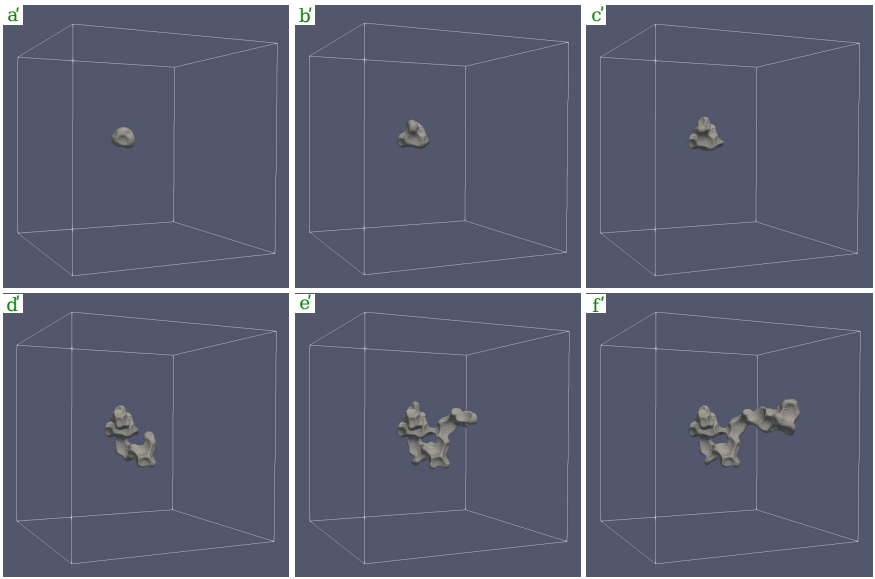}
\caption{The gas phase distribution for a drainage simulation of the 3D granular packing model with a small $S_r$ decrement ($1/10^{th}$ of the decrement used for the simulation shown in Figure \ref{SWCC_3D}). The suction and $S_r$ values of these points have been labeled in Figure \ref{increment_comparison}.}
\label{gas_cluster_2}
\end{figure*}

A question that arises here is why the drainage curve for the 3D model, shown in Figure \ref{SWCC_3D}, lacks noticeable pore emptying patterns and is much smoother compared to the drainage curve for the 2D model, shown in Figure \ref{SWCC_2D}. The answer to the first part of the question, regarding the lack of patterns, lies in the $S_r$ decrement used. A relatively large $S_r$ decrement during drainage results in emptying of many pores during a single drainage step, causing the patterns to disappear. See for instance how the gas cluster in Figure \ref{gas_cluster} quickly branches out to many different pores in the first 6 steps of drainage (the suction-$S_r$ for these points are labeled in Figure \ref{increment_comparison}). In order to see the patterns we saw in 2D, the $S_r$ decrement has to be reduced significantly. In Figure \ref{increment_comparison}, we plot the original drainage SWCC, which has an average $S_r$ decrement of 1.5$\%$, alongside a drainage SWCC simulated with an average $S_r$ decrement of 0.15$\%$. We see that similar patterns to the 2D case becomes obvious for the curve with a smaller decrement. Comparing the labeled points of the smaller increment curve (dashed green) in Figure \ref{increment_comparison} to the visualization of the gas cluster in Figure \ref{gas_cluster_2} shows that the source of the pattern is consistent with what we had concluded in 2D; the suction increases as the gas cluster extends into the throats (Point a$^\prime$ to b$^\prime$ to c$^\prime$), and then drops when the gas cluster enters a new chamber (Point d$^\prime$). Since simulations with a smaller decrement require a very long computation time, and given that the suction values of the two curves generated with different $S_r$ decrements are only slightly different, we have simulated the curve with the lower decrement only down to $S_r$ of 75$\%$. Regarding the smoothness of the curve, we believe that the main reason for this observation is that our 3D model is much less constrained than our 2D model due to the difference in model sizes as well as the additional degree of freedom in 3D. In our small 2D model, there were very limited pore options available for the gas cluster to extend to, while there are many more pore options available in the larger 3D model. Although we can still see oscillations happening in 3D if the drainage is performed very slowly, as shown in Figure \ref{increment_comparison}, the amplitude of the oscillations is much smaller than what we saw in 2D. The oscillation amplitudes become even smaller as $S_r$ decreases because the number of pore spaces the gas can expand into increases as the gas branches out to more pores, as seen in Figure \ref{gas_cluster}.

\begin{figure*}[tbh]
\centering
\includegraphics[width=1\textwidth]{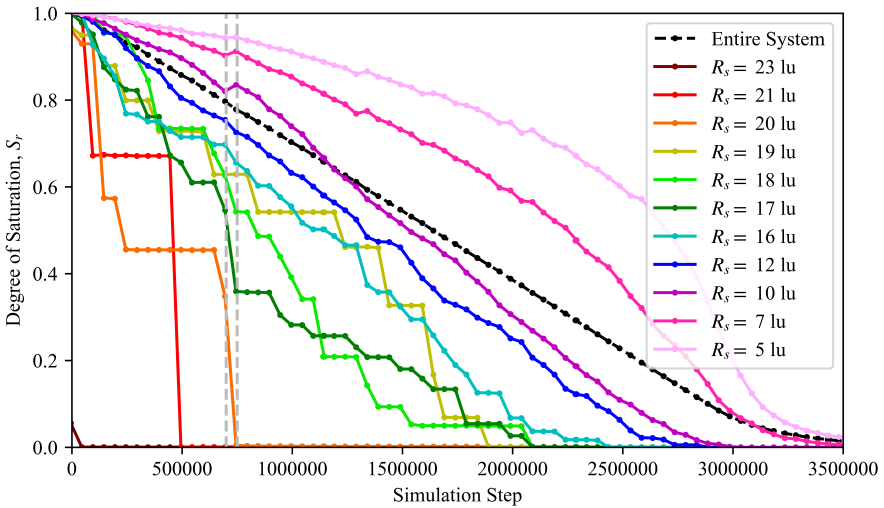}
\caption{Emptying of chambers with different sizes for the 3D granular packing model. $R_s$ refers to the radius of the inscribed sphere to each chamber and represents the chamber size. All chambers from the largest down to size 16 are shown. Below size 16, only a few selected sizes are shown. Refer to the text for explanation about the zone between the dashed gray lines.}
\label{desturation}
\end{figure*}

For the 2D model we saw that chambers do not empty in the order of descending size. To verify if this is also true for the 3D model, we measure the chamber sizes in our 3D model by measuring the radius of the largest sphere that can be inscribed in each chamber \cite{Sweeney2003}, $R_s$, and monitor the desaturation of chambers with different sizes. In Figure \ref{desturation}, we plot the degree of saturation of chambers grouped according to their size. At the start of the simulation, a gas cluster is residing in the largest chamber, with size 23, and therefore, this chamber is almost desaturated. The first chamber to empty as a result of drainage is one of the two chambers with the next largest size, 21 (there is no size 22 in the system), that empties during the second drainage step. Before the next chamber with size 21 empties at simulation step 500,000, the chambers with size 20 take lead and desaturate down to about 40$\%$, and also some other smaller chambers with sizes between 16 and 19 desaturate down to about 75$\%$. This is simply because the second chamber with size 21 is further away from the expanding gas cluster, compared to some other smaller chambers. Also notice how most chambers with size 17 empty before chambers with size 18, and similarly for chamber size 18 versus 19, again due to accessibility of the gas cluster. From chamber size 16 to 10, although not shown in Figure \ref{desturation} to avoid cluttering, we observe a similar trend, where there is no particular order in terms of size for the emptying of the chambers. Our analysis shows that chamber sizes below 10 are mainly extensions of larger chambers rather than being independent chambers surrounded by throats. Therefore, these chambers do in fact desaturate in order of their size as the gas cluster pushes into the throats. An interesting observation here is that the smaller chambers do not monotonically desaturate. For instance, see the zone between the dashed gray lines in Figure \ref{desturation}. During this drainage step a number of major chambers empty, and, as a result, the menisci can take larger radii of curvature; therefore, the menisci in the smaller chambers advance and fill up those chambers. In general, if we consider chambers with size greater than 10 as major chambers, we see that the same observation we made in 2D applies here, where major chambers do not empty according to their size.

\subsubsection{The pore filling process}\label{3d_pore_filling}

\begin{figure*}[tbh]
\centering
\includegraphics[width=1\textwidth]{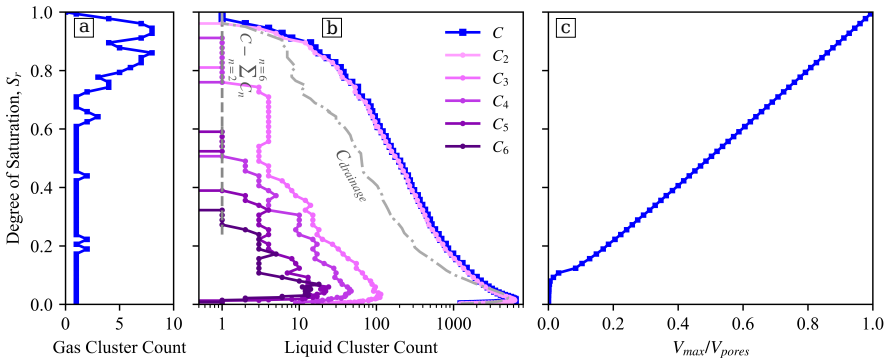}
\caption{Gas and liquid phase distribution statistics for the injection simulation of the 3D granular packing model. $C_{drainage}$ is equivalent to $C$ in Figure \ref{clusters_drain}.
}
\label{clusters_inject}
\end{figure*}

We use the gas and liquid cluster statistics plotted in Figure \ref{clusters_inject} to describe the pore filling process. In Figure \ref{clusters_inject}a, we see that, from the start of injection up to an $S_r$ of 60$\%$, there is mainly one gas cluster in the system, however, the number of gas clusters starts increasing at $S_r$ of about 60$\%$ and decreases again at $S_r$ above 95$\%$. This observation is consistent with the 2D model behavior, where the gas phase was split into isolated zones as a result of liquid bridges forming at empty throats. In Figure \ref{clusters_inject}b, we plot the liquid clusters count for different cluster orders during injection. We also plot the total number of liquid clusters during drainage for comparison. We see that the total number of liquid clusters during injection ($C$) is larger than the total number of liquid clusters during drainage ($C_{drainage}$). Particularly, the peak total number of liquid clusters of order 2 is 30$\%$ higher for injection compared to drainage, showing that during injection many new liquid bridges appear. Also clusters of higher order, for instance orders 3 to 6 shown in Figure \ref{clusters_inject}b, are present in the system for a wider range of $S_r$ during injection compared to drainage. Subtracting cluster counts of order 2 to 6 from the total cluster count shows that there is another cluster in the system, and our analysis reveals that this cluster has an order about 1068, meaning that it is connected to most of the grains. Figure \ref{clusters_inject}c shows that this single large liquid cluster starts forming at $S_r$ about 9$\%$ and constitutes the majority of the liquid volume from that point forward.

\begin{figure*}[tbh]
\centering
\includegraphics[width=1\textwidth]{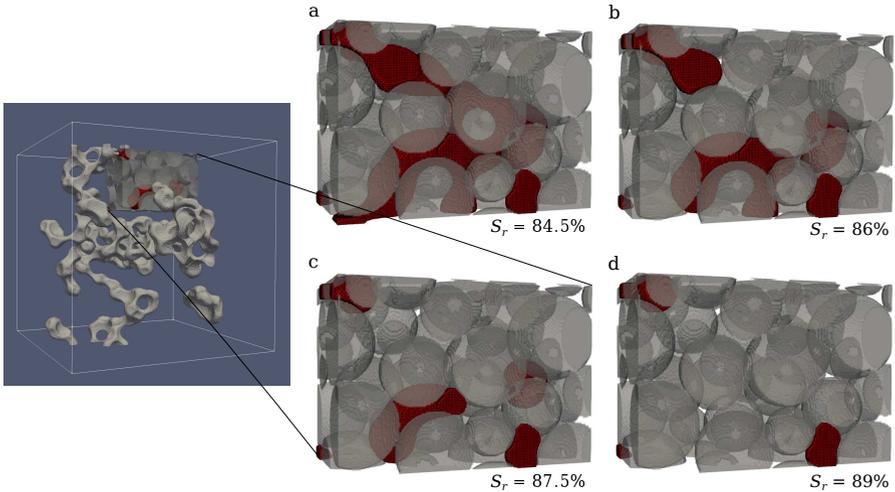}
\caption{The gas phase distribution in a subdomain of the 3D granular packing model, at different $S_r$ levels during the injection simulation. Gas clusters are shown in red for easier visualization.}
\label{gas_cluster_3}
\end{figure*}

Based on the observations above, we can summarize the pore filling process as follows. In terms of the gas phase distribution, initially there is only a single gas cluster and the gas phase is continuous, however, at $S_r$ above 60$\%$ the number of gas clusters increases, due to liquid bridges forming at empty throats and filling of smaller chambers that disconnect the gas zone. See for instance in Figure \ref{gas_cluster_3}, how the gas cluster, shown in red, splits into three separate clusters from injection step a to b. As $S_r$ increases further, the gas cluster count goes back down as the liquid fills in the chambers and gas clusters collapse. See for instance in Figure \ref{gas_cluster_3}, how two of the gas clusters vanish from injection step c to d. 

\begin{figure*}[h]
\centering
\includegraphics[width=1\textwidth]{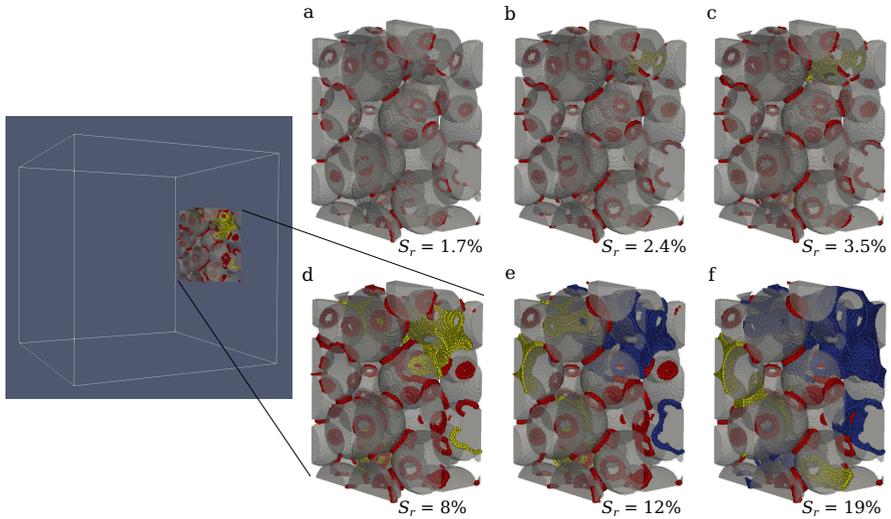}
\caption{The liquid phase distribution in a subdomain of the 3D granular packing model, at different $S_r$ levels during the injection simulation. The single large liquid cluster is shown in blue, liquid bridges are shown in red, and all other clusters are shown in yellow.}
\label{liq_cluster_2}
\end{figure*}

\begin{figure*}[!h]
\centering
\includegraphics[width=1\textwidth]{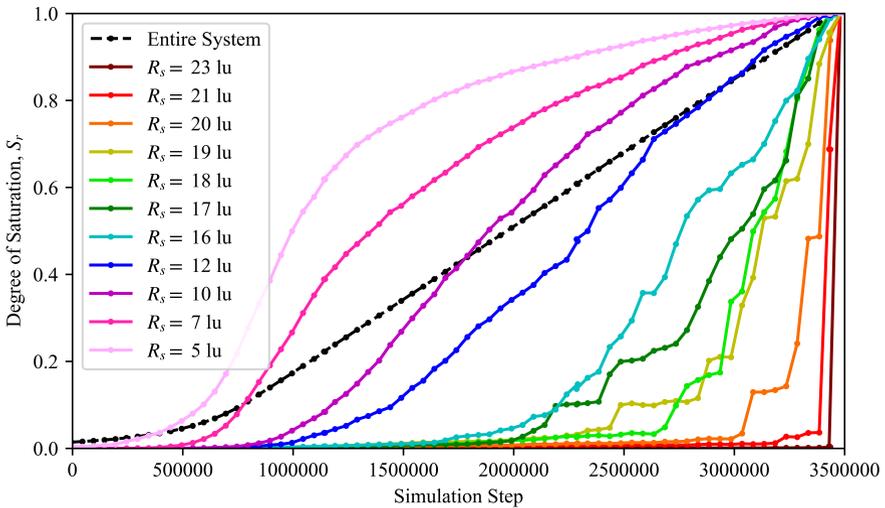}
\caption{Filling of chambers with different sizes for the 3D granular packing model. $R_s$ refers to the radius of the inscribed sphere to each chamber and represents the chamber size. All chambers from the largest down to size 16 are shown. Below size 16, only a few selected sizes are shown.}
\label{saturation}
\end{figure*}

In terms of the liquid phase distribution, the liquid cluster count initially increases as new bridges are formed, and it reaches a maximum value at $S_r$ about 1$\%$. At $S_r$ above 1$\%$, the number of liquid clusters decreases as a result of clusters joining together, as shown in Figure \ref{liq_cluster_2}. Particularly, at $S_r$ between 9$\%$ and 12$\%$ the liquid clusters come together to form a single large cluster that is in constant contact with most of the grains. This cluster is shown in blue in Figure \ref{liq_cluster_2}. At $S_r$ above 12$\%$, the majority of the liquid volume is taken by the single large liquid cluster, but there are also a high number of liquid bridges. At $S_r$ above 98$\%$ all the liquid bridges join the large cluster, and there is only one large cluster in the system that immerses all the grains. Figure \ref{saturation} shows that during pore filling, unlike pore emptying, the chambers fill in order of increasing size. The pore filling process we see here is consistent with what we had previously identified in 2D.

\subsubsection{Comparison of pore emptying and pore filling processes: source of hysteresis}\label{3d_comparison}

\begin{figure}[tbhp]
\centering
\includegraphics[width=0.525\textwidth]{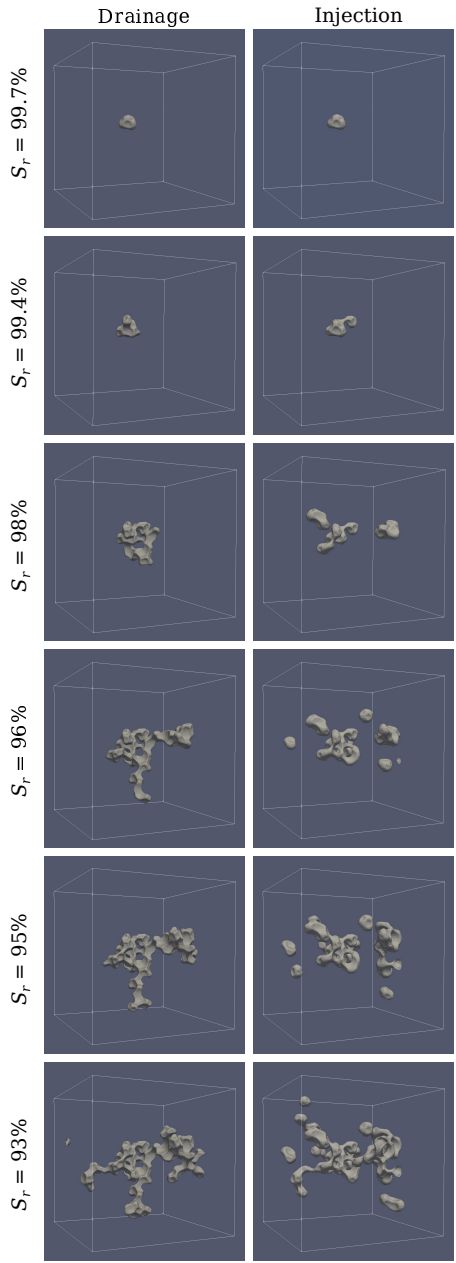}
\caption{Comparison of drainage and injection gas phase distributions for the 3D granular packing model.}
\label{comparison_3d}
\end{figure}

Using gas and liquid phase distribution statistics, and confirmed by visualizations, we verified that the pore emptying and pore filling processes we had identified in 2D are also true for 3D. During drainage only the existing gas cluster expands and no new gas clusters appear in the system, whereas during injection, liquid coalesces at empty throats, creating many liquid clusters that expand simultaneously. The constraint on where the gas cluster can expand to during drainage forces the menisci to take smaller radii of curvature, which is associated with the larger suction of the drainage SWCC. The smaller meniscus radii of curvature during drainage and the higher number of gas clusters during injection, at $S_r$ levels below 99.7$\%$, can be visualized in Figure \ref{comparison_3d}.

\begin{figure*}[tbh]
\centering
\includegraphics[width=0.95\textwidth]{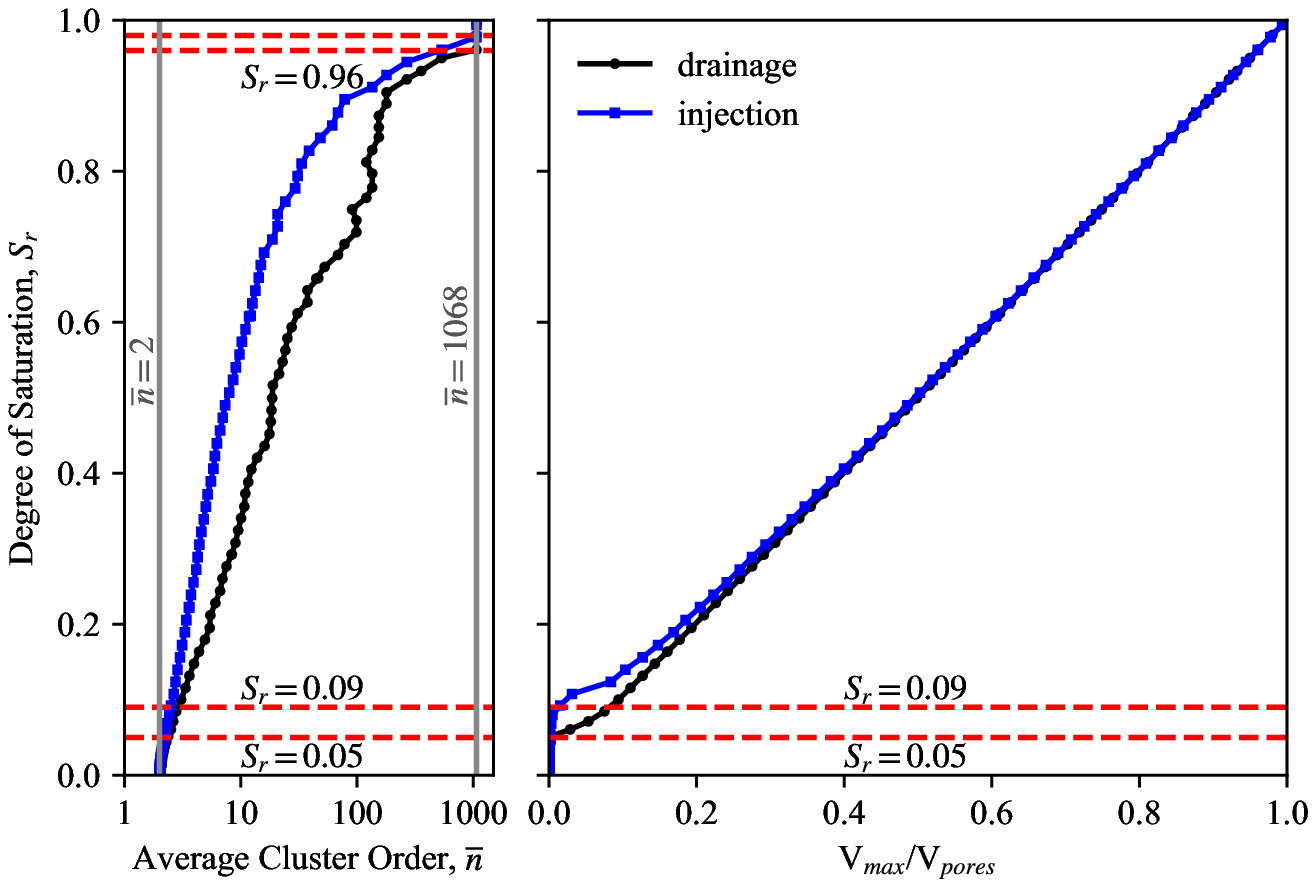}
\caption{Cluster statistics used for identifying the liquid content states. The average cluster order is calculated as $\overline{n}=\sum_{n=2}^{n=1068} nC_n/C$. $V_{max}$ is the volume of the largest liquid cluster in the system, and $V_{pores}$  is the total volume of the pores.
}
\label{states_3d}
\end{figure*}

\begin{figure*}[tbh]
\centering
\includegraphics[width=0.475\textwidth]{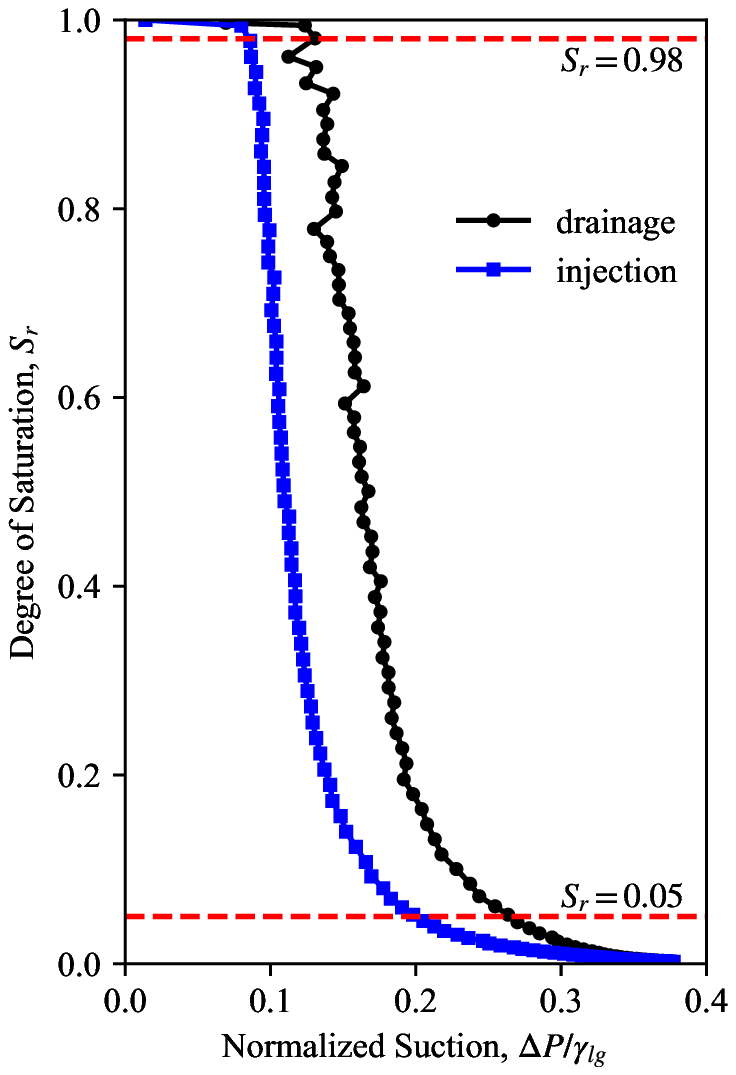}
\caption{Soil-water characteristic curves for drainage and injection simulations of the 3D granular packing model with the addition of liquid-content-states indicators. The red dashed lines separate the different liquid content states.}
\label{SWCC_3D_2}
\end{figure*}

For the 2D model the drainage and injection SWCCs matched in the pendular and capillary states. To check if this observation holds for the 3D model, we need to first identify the different liquid content states. We use the average cluster order as well as the $V_{max}/V_{pores}$ value, which we previously introduced, to identify these states. In Figure \ref{states_3d}, we plot these two quantities for both drainage and injection. The pendular state is where the liquid is mainly in the form of bridges between grains, therefore, the average cluster order should be about 2. Also, since the liquid bridges have a very small volume compared to total pore space, $V_{max}/V_{pores}$ should be about zero. Based on these two criteria, we find the pendular state to be below $S_r$ = 5$\%$ for drainage and below $S_r$ = 9$\%$ for injection. The capillary state is where all grains are immersed in liquid, therefore, the average cluster order should be equal to the number of grains, 1068 in our case. Based on this criterion, we find the capillary state for the drainage and injection to be above $S_r$ = 96$\%$ and 98$\%$, respectively. If we now consider the ranges of $S_r$ over which both drainage and injection have the same liquid content state, that is $S_r$ above 98$\%$ for the capillary state and $S_r$ below 5$\%$ for the pendular state, we see in Figure \ref{SWCC_3D_2} that the drainage and injection SWCCs surprisingly do not match in these ranges of $S_r$. In fact, it appears that for the 3D model, the hysteresis expands over the entire range of $S_r$.

Although we verified that the source of hysteresis for the 3D model is identical to what we had found for the 2D model, there seems to be a difference in the $S_r$ range where hysteresis exists for the 3D model versus the 2D. Our findings have shown that the suction during drainage and injection only match when the liquid and gas phase distributions are similar. This is because suction is a function of the menisci radius of curvature and, for a given $S_r$, the radius of curvature is a function of the distribution of the phases. Based on what we have learned about the differences between the pore emptying and the pore filling processes, the only times that the phase distributions can be similar is either when there are only liquid bridges in the system or when there is only a single gas cluster in a single chamber. 

For the 2D model, the last chamber emptied at an $S_r$ above 5$\%$ and the first chamber did not fill until $S_r$ of about 12$\%$, therefore, below 5$\%$ there were only liquid bridges in the system and the gas and liquid phase distributions matched for injection and drainage. However, in the 3D model, because there are many more chambers available with a wide range of sizes, emptying of chambers continues until full desaturation, and filling of chambers starts immediately with the start of injection, therefore, the liquid and gas phase distributions are different even at very low $S_r$ and the SWCCs do not match. This argument is supported by the fact that for the 3D model, clusters of order higher than two are present in the system even at very low $S_r$ (See Figure \ref{clusters_drain}b and Figure \ref{clusters_inject}b), and while the average cluster order converges to two at low $S_r$, it does not become exactly two until $S_r$ is about zero (see Figure \ref{states_3d}), showing that a true pendular state where there are only bridges in the system does not occur for the 3D model. 

As for the capillary state, in Figure \ref{comparison_3d} we see that at $S_r$ of 98$\%$ which we identified as the start of the capillary state, there are four gas clusters during injection while there is only one during drainage, and therefore, the phase distributions do not match. Even at $S_r$ of 99.4$\%$ where there is a single gas cluster in both injection and drainage, the shape of the clusters are entirely different because the gas cluster for injection is connecting two chambers while the gas cluster for drainage is only inside a single chamber (since the AEV of the neighboring chamber has not been reached yet). Only at a very high $S_r$ of 99.7$\%$, where there is a single gas cluster inside a single chamber, we see that the phase distributions become similar and the suction values for injection and drainage match. We observed a similar behavior for the 2D model, however, because the model size was small, the formation of a single gas cluster inside a single chamber occurred at $S_r$ above 66$\%$ and coincided with the capillary state, whereas, for the 3D model, a single gas cluster inside a single chamber has a negligible volume compared to the entire pore space and corresponds to an $S_r$ of about 100$\%$. Therefore, we conclude that, for large models, SWCC hysteresis occurs over the entire range of $S_r$, regardless of liquid content state.

\section{Summary and conclusion}\label{conclusion}

We investigate the underlying source of Soil-Water Characteristic Curve (SWCC) hysteresis by means of pore-scale numerical simulation, using the multiphase Lattice Boltzmann Method (LBM). Starting from a simple configuration consisting of a liquid bridge between two solid plates, we show how suction appears in a multiphase system and introduce the governing Young-Laplace equation. We then show that for menisci between planer surfaces the radius of curvature, and hence suction, is constant regardless of the degree of saturation ($S_r$), while for menisci between non-planar surfaces,  the radius of curvature, hence suction, depends of the $S_r$. We demonstrate the latter by simulating a liquid bridge between two disks, and showing that the suction monotonically increases/decreases with the decrease/increase of $S_r$. We also show that there is no suction-$S_r$ hysteresis among the drainage and injection paths for the liquid bridge between two disks. We then move on to a small 2D granular packing consisting of 15 grains, for which we observe SWCC hysteresis. We investigate the source of  the hysteresis by comparing the pore emptying and pore filling processes. Finally, we simulate a larger 3D granular packing with 1068 grains, and verify that the source of hysteresis identified using the 2D model still applies.

We find the source of hysteresis in our model in the difference between the pore filling and the pore emptying processes. The pore filling process can be thought of as the expansion of the liquid zone during injection. When slowly injecting liquid in the system, new liquid zones appear in the gas zone in the form of liquid bridges, due to capillary condensation. Therefore, during pore filling, many liquid bridges expand simultaneously, and join together to fill the pores from the smallest to largest. The pore emptying process can be thought of as the expansion of the gas zone during drainage. When slowly draining liquid out of the system, new gas zones do not appear in the liquid zone because that would require an extremely high suction. Therefore, during pore emptying, only the existing gas cluster expands and the pores are emptied according to their adjacency to the gas cluster and not necessary from the largest to smallest. As a result, the expansion of the gas cluster during pore emptying is constrained by the size of the pore openings surrounding it, forcing the menisci to take the small radius of curvature required to enter those pores, and therefore, making the suction during drainage larger than the suction during injection, at a given $S_r$.

\subsection*{Declarations}

\textbf{Conflict of Interest} The authors declare that they have no conflict of interest.


\bibliography{hysteresis}


\end{document}